\documentclass[preprint, 11pt]{revtex4}

\usepackage[utf8]{inputenc}
\usepackage[svgnames, dvipsnames]{xcolor}
\usepackage{amsfonts, amssymb, amsmath, stmaryrd, mathtools, mathrsfs, abraces, bm}

\usepackage{bbm}
\DeclareMathAlphabet{\mathpzc}{OT1}{pzc}{m}{it}
\usepackage{graphicx}

\usepackage{algpseudocode}
\usepackage{algorithm}
\usepackage[toc,page]{appendix}
\usepackage[colorlinks=true, linkcolor=MediumBlue, citecolor=blue,
urlcolor=blue]{hyperref}

\makeatletter
\newcommand{\specialcell}[1]{\ifmeasuring@#1\else\omit$\displaystyle#1$\ignorespaces\fi}
\makeatother

\newtheorem{theorem}{Theorem}

\newtheorem{lemma}{Lemma}

\newcommand*{\1}{\mathbbm{1}}
\newcommand*{\dd}{\mathrm{d}}
\newcommand*{\eq}[1]{(\ref{eq:#1})}

\newcommand*{\set}{\Omega}
\newcommand*{\setind}{\llbracket 1, N \rrbracket}
\newcommand*{\setvar}{\mathcal{V}}
\newcommand*{\var}{v}

\newcommand*{\pdmpX}{X}
\newcommand*{\pdmpVar}{V}
\newcommand*{\pdmpx}{x}

\begin{document}

\title{PDMP characterisation of event-chain Monte Carlo algorithms for particle systems}

\author{Athina Monemvassitis}
\email{athina.monemvassitis@uca.fr}
\author{Arnaud Guillin}
\email{arnaud.guillin@uca.fr}
\author{Manon Michel}
\email{manon.michel@uca.fr}
\affiliation{Laboratoire de Math\'ematiques Blaise Pascal UMR 6620, CNRS, Université Clermont-Auvergne, Aubière, France.}

\begin{abstract}
  Monte Carlo simulations of systems of particles such as hard spheres
  or soft spheres with singular kernels can display around a phase
  transition prohibitively long convergence times when using
  traditional Hasting-Metropolis reversible schemes. Efficient
  algorithms known as event-chain Monte Carlo were then developed to
  reach necessary accelerations. They are based on non-reversible
  continuous-time Markov processes. Proving invariance and ergodicity
  for such schemes cannot be done as for discrete-time schemes and a
  theoretical framework to do so was lacking, impeding the
  generalisation of ECMC algorithms to more sophisticated systems or
  processes. In this work, we characterize the Markov processes
  generated in ECMC as piecewise deterministic Markov processes. It first
  allows us to propose more general schemes, for instance regarding
  the direction refreshment. We then prove the invariance of the correct
  stationary distribution. Finally, we show the ergodicity of the processes in soft-
  and hard-sphere systems, with a density condition for the latter.

\end{abstract}

\maketitle

\section{Introduction}

Since the introduction of the Markov-chain Monte Carlo method (MCMC)
\cite{Metropolis_1953}, continuous particle systems evolving according
to pairwise and central interactions have filled the part of both an
important scientific motivation in itself and a formidable testbed and
accelerator for MCMC development. Indeed, despite the simplicity of
the interactions (e.g. hardcore, Lennard-Jones or powerlaw-decaying
repulsions), the behaviours displayed by these systems are rich. They
are however described by high-dimensional integrals, whose analytical
resolution remains ouf of reach. Their MCMC numerical evaluation
by a discrete sum over a large collection of system
configurations has thus been an important focus of the computational
physics \cite{Alder_1962,Jaster_1999,Bernard_2011} as well as of the
probability community \cite{Diaconis_2011} for example. MCMC methods
rely on Markov processes to produce system configurations in a random
sequence. Such sequence should preferrably exhibit the least
correlations possible between two succeeding states, in order to
minimise the asymptotic error on the integral estimation
\cite{Janke_2002}. Typically, reversible Markov processes rely on move
rejections for correctness and display a diffusive dynamics
\cite{Levin_2017}. They then show important correlations, which can
furthermore be greatly increased in presence of critical slowing down
phenoma at phase transitions \cite{Hohenberg_1977}. The development of
MCMC methods which can break free from any diffusive behavior and slow
physical time scales has thus been instrumental to reach system sizes
big enough to control finite-size effects, as achieved by the
rejection-free cluster algorithms in lattice-spin sytems
\cite{Swendsen_1987,Wolff_1989}.

In continuous particle systems, the lack of a natural involution
symmetry, necessary to ensure the reversibility and correctness of
cluster methods, has led to the development of non-reversible MCMC
methods\cite{Turitsyn_2011,Peters_2012}, among which Event-Chain Monte
Carlo (ECMC) \cite{Bernard_2009,Michel_2014}. By producing persistent
global moves forming up a sequence of ballistic trajectories of single
particles, the accelerations brought by ECMC allowed to resolve the
heavily-debated question of the scenario for the melting of
bidimensional hard disks \cite{Widom_1987,Bernard_2011} and to
investigate the same phenomenon with power-law decaying interactions
\cite{Kapfer_2015}. These successes have thus motivated the
application of ECMC algorithms to other classic systems in statistical
physics (polymers \cite{Kampmann_2015}, continuous spins
\cite{Michel_2015,Nishikawa_2015}) and started the development of
generalized and accelerated ECMC variant (e.g. generalisation to
n-body interactions \cite{Harland_2017}, Forward ECMC
\cite{Michel_2020}, Newtonian ECMC
\cite{Klement_2019,Hoellmer_2021}). Stemming from this research line
in statistical physics, similar non-reversible Markov processes as the
ones used in ECMC were applied to sampling problems in statistical
Bayesian inference \cite{Bouchard_2018,Bierkens_2019_zz} and were
mathematically framed as Piecewise Deterministic Markov Processes
(PDMP) \cite{Davis_1984,Davis_1993}, leading to the general name of
Piecewise Deterministic Monte Carlo (PDMC). Piecewise deterministic
Markov processes have already been studied in queuing theory
\cite{Davis_1993}, biology \cite{Radulescu_2007,Yvinec_2015} and
physics, see for example \cite{M2015} for a review of a variety of
PDMP applications. Now under focus in Bayesian statistics, efforts
have been made towards a rigorous analysis of the properties of
invariance, ergodicity and convergence of the processes found in PDMC
\cite{Durmus_2018,Fearnhead_2018,Bierkens_2019_zz}.

In this work, we push this analytical effort and characterize the PDMP
generated by ECMC in bidimensional disk systems. This allows to prove
the invariance of the commonly-used schemes and gives a framework to
do so for future algorithmic upgrades or applications. In particular,
we show how the refreshment of the direction of the ballistic
trajectories can be described by a boundary effect instead of a
Poisson process, as commonly found in PDMC in the statistics
literature. Doing so reflects better the algorithmic implementations
as done in physics, as fixed-time refreshment schemes can greatly ease
the computations. Furthermore, this allows us to give conditions on
valid refreshment schemes, opening up to a larger choices of
strategies than the exponential refreshment generated by a Poisson
process or the uniform one as obtained by a fixed-time
scheme. Finally, we study the ergodicity property of those processes
for soft and hard disks. We use the Harris recurrence theory
\cite{Meyn_1993,Tweedie_1994,Meyn_2012} starting from the uniform
continuity property of the process \cite{Bakhtin_2012,Benaim_2015} and
adapting the proof done in \cite{Bierkens_2019}. In presence of
hardcore repulsions, even at low densities and for reversible Markov
chains, proving ergodicity is a hard problem, first starting with a
usually simple task such as showing irreducibility. Here, we follow the
approach used in \cite{Diaconis_2011} for a reversible Markov chain,
so as to show connectivity of the state space, but obtain ergodicity
for higher densities.

In the present paper, we start by introducing the ECMC method and its
implementation in disk systems in Section \ref{sec:ecmc}. We then
present the PDMP characterisation of ECMC and in particular of its
refreshment strategy in Section \ref{sec:pdmp}. This allows us to
derive the invariance in Section \ref{sec:inv} through a generator
characterization and the ergodicity in Section \ref{sec:ergo}.

\section{Event-chain Monte Carlo for soft- and hard-sphere systems}
\label{sec:ecmc}
\subsection{Soft- and hard-sphere systems}

Event-chain Monte Carlo algorithms were first developed for hard- and
soft-sphere systems in a periodic bidimensional box
\cite{Bernard_2009,Michel_2014}. Both systems share a common
description, and henceforth we will only use subscripts $S$ or $H$
when making statements restricted to respectively the soft- or
hard-sphere systems.

For both systems, a configuration of the $N\in\mathbb{N}$ spheres of
radius $\sigma\in\mathbb{R}_+^*$ in a periodic box of length
$L\in\mathbb{R}_+^*$ is completely characterized by the sphere
positions 
$x = (x_i=(x_{i,0},x_{i,1}))_{i\in\setind}\in\Omega(N)\subset
(\mathbb{R}/L\mathbb{Z})^{2N}$. The configuration set $\Omega(N)$ of valid
configurations is an open set, completely defined according to the excluded minimal
pairwise distance $d_{\text{pair}}\in\mathbb{R}_+$, as,
\begin{equation}
  \Omega(N)=\{x\in(\mathbb{R}/L\mathbb{Z})^{2N};\forall  (i,j)\in\setind^2,i\neq j, (x_i,x_j)\in\Omega_{\text{pair}}\},
  \label{eq;set}
\end{equation}
with $\set_{\text{pair}}$ the open set of valid pair of positions,
\begin{equation}
  \Omega_{\text{pair}}=\{(x,x')\in(\mathbb{R}/L\mathbb{Z})^{4};d(x,x')>d_{\text{pair}}\}
  \label{eq:pair}
  \end{equation}
  where $d$ is the $L$-periodic distance
  $d:(\mathbb{R}/L\mathbb{Z})^4\to\left[0,\frac{L}{\sqrt 2}\right]$,
  which corresponds for any pair of positions
  $(x=(x_0,x_1),x'=(x'_0,x'_1))\in (\mathbb{R}/L\mathbb{Z})^4$ to the
  minimal distance between all their periodic copies,
\begin{equation}
 d(x,x') = \sqrt{\sum_{k=0}^1\min(|x_{k}-x'_{k}|, L - |x_{k}-x'_{k}|)^2}.
  \label{eq:d}
\end{equation}
The configuration set $\set(N)$ presents the boundary,
\[ \partial \set(N)=\{x\in(\mathbb{R}/L\mathbb{Z})^{2N};\forall
  (i,j)\in\setind^2,i\neq j, (x_i,x_j)\in\Omega_{\text{pair}}\cup\partial\Omega_{\text{pair}}\}\setminus\Omega,  
\]
with \[\partial \Omega_{\text{pair}}=\{(x,x')\in(\mathbb{R}/L\mathbb{Z})^4;d(x,x')=d_{\text{pair}}\}.\]
From now on, we will drop the dependence in $N$ of $\Omega(N)$ for simplicity, when there is no possible confusion on the number $N$ of spheres involved.

Sphere configurations then follow at equilibrium a Boltzmann distribution,
\begin{equation}
   \forall x \in \Omega, \pi(x)\propto \prod_{1\leq i<j\leq N} \exp[-\beta u(d(x_i,x_j))],
  \label{eq:pi}
\end{equation}
with $\beta$ the inverse temperature, set to $1$ in the following, and
$u: ]d_{\text{pair}}, +\infty[\to \mathbb{R}_+$ a continuous and piecewise
differentiable function, which codes for the potential energy arising
from the pairwise interactions and depending only on the pairwise
periodic distance. We define by continuity the extended distribution
$\tilde{\pi}$,
\begin{equation}
  \begin{array}{ll}
  \forall x \in \Omega\cup\partial\Omega, \tilde{\pi}(x) \propto&
 \lim_{r\to d_{\text{pair}}^+}\exp\left[-\beta u(r)\sum_{1\leq i<j\leq N}\1_{\partial \Omega_{\text{pair}}}((x_i,x_j))\right]\\
&   \times \exp\left(-\beta \sum_{1\leq i<j\leq N}u(d(x_i,x_j))(1-\1_{\partial \Omega_{\text{pair}}}((x_i,x_j)))\right]
    \end{array}
  \label{eq:tilde_pi}
\end{equation}

Explicitely, for soft spheres, 
\begin{equation}
  \left\{
  \begin{array}{l}
  \Omega_S = (\mathbb{R}/L\mathbb{Z})^{2N}\setminus\{x\in(\mathbb{R}/L\mathbb{Z})^{2N}~;
  \exists{(i,j)}\in\setind^2, i\neq j,
             d(x_i,x_j)=0\}\\
  d_{S,\text{pair}}=0\\
    \Omega_{S,\text{pair}}=\{(x,x')\in(\mathbb{R}/L\mathbb{Z})^{2};d(x,x')>0\},
      \end{array}\right.
 \label{eq:setS}
\end{equation}
and the interactions are ruled for any pair interdistance $r\in\mathbb{R}^*_+$,by
  \begin{equation}
    u_S(r) =\left\{
      \begin{array}{ll}
        \left(\frac{\sigma}{r}\right)^\gamma - \left(\frac{\sigma}{r_c}\right)^\gamma
        &\text{for $r \leq r_c$}\\
        0&\text{otherwise}\\
        \end{array}\right.,
    \label{eq:u}
    \end{equation}
    with $r_c\in \left]0,\frac{L}{2}\right]$ a cut-off
    length. It leads to the soft-sphere equilibrium distribution
    $\pi_S$ and corresponding extended one $\tilde{\pi}_S$,
    \begin{equation}
      \left\{
        \begin{array}{ll}
      \forall x \in \Omega_S,& \pi_S(x)\propto \prod_{1\leq i<j\leq N} \exp[-\beta u_S(d(x_i,x_j))]\\   
   \forall x \in\Omega_S\cup \partial\Omega_S,& \tilde{\pi}_S(x) = \1_{\Omega}(x)\pi(x).
 \end{array} \right.
    \label{eq:pisoft}
\end{equation}  

Now, for hard spheres,
\begin{equation}
  \left\{
  \begin{array}{l}
  \set_H = (\mathbb{R}/L\mathbb{Z})^{2N}\setminus\{x\in(\mathbb{R}/L\mathbb{Z})^{2N};\exists(i,j)\in\setind^2, i\neq j, d(x_i,x_j)\leq 2\sigma\}\\
  d_{H,\text{pair}} = 2\sigma\\
    \Omega_{H,\text{pair}}=\{(x,x')\in(\mathbb{R}/L\mathbb{Z})^{2};d(x,x')> 2\sigma\}.
      \end{array}\right.
  \label{eq:setH}
\end{equation}
For hard-sphere systems, there is no interaction apart from the
hard-core repulsions which are already encoded in the non-zero
$d_{H,\text{pair}}$, so that we have for any $r\in ]2\sigma, +\infty[$,
  \begin{equation}
    u_H(r) = 0,
    \label{eq:u}
    \end{equation}    
and the hard-sphere equilibrium distribution $\pi_H$ (resp. the extended distribution $\tilde{\pi}_H$) is then the uniform one on $\set_H$ (resp. on $\set_H\cup\partial\set_H$),

\begin{equation}
  \left\{
    \begin{array}{ll}
  \forall x \in \Omega_H, &\pi_{H}(x)\propto \mathbbm{1}_{\Omega_H}(x)\\
      \forall x \in \Omega_H\cup\partial\Omega_H,& \tilde{\pi}_{H}(x)\propto \mathbbm{1}_{\Omega_H\cup\partial\Omega_H}(x).
        \end{array}\right.
  \label{eq:pihardHeaviside}
  \end{equation}

\subsection{Event-chain Monte Carlo}
\label{sec:ecmc_pres}
Sampling configurations in a set $\Omega$ according to a target
probability distribution $\pi$ is achieved in a MCMC method through
the recursive application of a Markov kernel, denoted as $K$, such
that $\pi$ is left invariant, equivalently, such that the global balance
condition $\pi K=\pi$ is satisfied, i.e., 

\begin{equation}
  \int_{x'\in\Omega}\pi(\dd x')K(x',\dd x) = \int_{x'\in\Omega}\pi(\dd x)K(x,\dd x')=\pi(dx).
  \label{eq:GB}
  \end{equation}

  The seminal Metropolis algorithm \cite{Metropolis_1953}, was first
  developed for sampling from sphere systems by enforcing a sufficient
  condition to the global balance, the detailed balance, i.e.
  $\pi(\dd x')K(x',\dd x)=\pi(\dd x)K(x,\dd x')$ for every pair of
  configurations $(x,x')\in\Omega$, through the following choice for
  $K$,
  \begin{equation}
    K(x,\dd x') = q(x'|x)a(x'|x)\dd x' + \left(1-\int_{y \in \Omega}q(y|x)a(y|x)\dd y\right)\delta_{x=x'},
    \label{eq:Metro}
    \end{equation}
    where $q$ identifies with the proposal distribution, a common
    choice (e.g. as  studied in \cite{Diaconis_2011}) verifying
  $\int_{\Omega}q(x'|x)\dd x' = 1$ being,
  \begin{equation}
    q(x'| x)=\frac{1}{h^2\text{vol}(B_1)}\mathbbm{1}_{B_1}\left(\frac{x-x'}{h}\right),
    \label{eq:Metroprop}
  \end{equation}
  with $h\in]0,1]$ and $B_1$ the unit ball of $\mathbbm{R}^2$, and
  where $a$ identifies with the acceptance rate,
  \begin{equation}
    a(x'|x)=\min\left(1,\frac{\pi(x')}{\pi(x)}\right),
    \label{eq:Metroacc}
    \end{equation}
    generalized in \cite{Hastings_1970} to,
  \begin{equation}
    a(x'|x)=\min\left(1,\frac{q(x|x')}{q(x'|x)}\frac{\pi(x')}{\pi(x)}\right),
    \label{eq:Metroaccgen}
    \end{equation}
    for an asymmetric proposal distribution $q$.

    Event-chain Monte Carlo algorithms \cite{Bernard_2009,Michel_2014}
    were developed so that only the necessary condition of global
    balance is satisfied, while the detailed-balance one is
    broken. This is achieved by generating a non-reversible Markov
    process through the exploitation of the pairwise translational
    invariance or mirror symmetry for any pair of spheres
    $(i,j)\in \setind^2$, i.e.  $\nabla \cdot u(d(x_i,x_j))=0$ (resp.
    $\nabla \cdot H(d(x_i,x_j)) =0 $ in the limit of hard spheres
    systems, $H$ being the Heaviside function). ECMC schemes are now also
    generalized to systems presenting general n-body interactions by
    exploiting the global translational invariance
    ($\nabla \cdot u = 0$), see \cite{Harland_2017}.

    The initial introduction of these methods 
    relied on taking an infinitesimal limit and using continuous-time
    Markov processes, while extending the state $x$ to $(x,v)$, with
    $v$ an auxiliary variable, commonly referred to as the
    \emph{lifting} variable, following \cite{Diaconis_2000}. The
    purpose of such variable is to introduce persistence into the
    proposal distribution while ensuring the process remains
    Markovian. In bidimeniosal sphere systems, the state space $\set$
    is thus extended to $\set\times\setvar$ with
    $\setvar = \{(1,0),(0,1)\} \times\setind$, associated with the
    measure $\mu_\setvar = \mu_{\mathcal{D}}\otimes\mu_N$ where
    $\mu_{\mathcal{D}}$ and $\mu_N$ are the counting measures over
    $\mathcal{D}=\{(1,0),(0,1)\} $ and $\setind$ respectively. The
    Markov process should then be targetting as stationary
    distribution $\pi \otimes \mu_\setvar$. For any pair of
    configurations
    $(x,v=(e,i)),(x',v'=(e',i'))\in(\set\times\setvar)^2$ the
    acceptance is set to $1$ (i.e. $a((x',v')|(x,v))=1$). The
    auxiliary variable $v=(e,i)$ codes for proposing moves of the
    $i$-th sphere along the direction $e$, setting,
    \begin{equation}      
  \begin{aligned}
  &q_{\epsilon}((x',(e',i')))|(x,(e,i)))&\\
                  & =(1-r)\delta(e-e')\delta(i-i')\delta(x_i+\epsilon e - x'_i)\prod_{j\neq i}\delta(x_j-x'_j)\prod_{j\neq i}p_\epsilon(x_i,x_j,e)& \text{ (physical move)}\\
                  &+(1-r)\delta(e-e')(1-\delta(i-i'))\delta(x-x')(1 - p_\epsilon(x_i,x_{i'},e))\prod_{j\neq i, i'} p_\epsilon(x_i,x_j,e) &\text{(lifting move)}\\
                  &+r\delta(x-x')\mu_\setvar(\dd e',\dd i')&\text{(refreshment)}
  \end{aligned}
  \label{eq:fin_q_a}
\end{equation}
with $\epsilon\in \mathbb{R}_+^*$ the step magnitude, $0<r<1,$ the refreshment
probability and $p_\epsilon$ the \emph{factor} probability \cite{Michel_2014} for $(x_i,x_j)\in\Omega_{\text{pair}}$,
\begin{equation}
  p_\epsilon(x_i,x_j,e) = \min\left(1,\1_{\Omega_{\text{pair}}}((x_i+\epsilon e,x_j))\frac{\exp(-u(x_i+\epsilon e,x_j))}{\exp(-u(x_i,x_j))}\right),
\label{eq:pfact}
\end{equation}
which builds up a factorized variant of the usual Metropolis
probability \eq{Metroacc} and which also satisfies detailed balance in
a \emph{skewed} form, as
\begin{equation}\exp(-u(x_i,x_j)) \prod_{j\neq i }p_\epsilon(x_i,x_j,e)=\exp(-u(x_i+\epsilon e,x_j)) \prod_{j\neq i }p_\epsilon(x_i+\epsilon e,x_j,-e).
  \label{eq:fact_DB}
  \end{equation}
The idea behind this choice for the proposal distribution $q_\epsilon$
is to propose \emph{physical} moves (i.e. updates of $x$) by moving
the $i$-th sphere by $+\epsilon e$ until a \emph{rejection} triggered
by another sphere $i'$ through its pairwise interaction with $i$
occurs. Then, the physical move is replaced by a \emph{lifting} one
(i.e update of $v$ from $v=(e,i)$ to $v'=(e,i')$) and the $i'$-th sphere
is now the one being updated by $+\epsilon e$ increment. Eventually,
the proposal distribution includes a \emph{refreshment} term in order
to ensure irreducibility, which can also halt the persistent physical
moves to update $v$.

\begin{figure}[t] \centering
        \includegraphics[width=\textwidth]{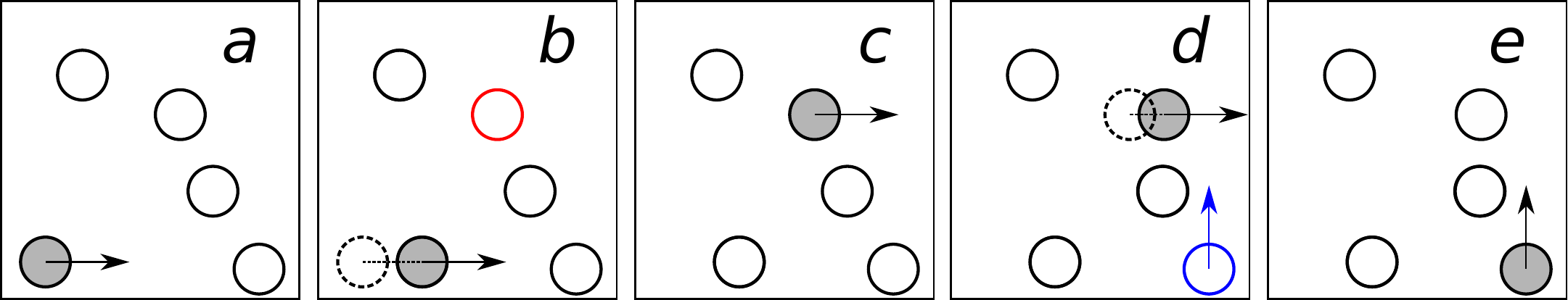}
        \caption{Illustration of moves produced from an ECMC algorithm
          in a system of soft spheres. The grey sphere, set by the
          label $i$, is updated along the direction $e=(1,0)$
          (physical move, \emph{a}) until a first event occurs, here
          with the $j$-th sphere (in red, \emph{b}). The label then is
          updated from $i$ to $j$ (lifting move, \emph{c}) and it is
          the $j$-th sphere which is now being updated along $e$
          before being stopped at a refreshment time (\emph{d}), where
          a new label and direction (in blue, \emph{d}) is resampled
          and from which a new chain of physical and lifting moves is
          produced (\emph{e}).}
        \label{fig:evolution-pdmp}
      \end{figure}

Nonetheless, in spite of the property \eq{fact_DB} of the factor probabiliy
$p_\epsilon$, the proposal distribution \eq{fin_q_a} does not define a
valid MCMC scheme for finite $\epsilon$ since \emph{rejections} of the
physical move from multiple pairs at once are not accounted for, i.e.,
\begin{equation}
\begin{split}\int\dd x'\dd e'\dd i' q_\epsilon((x',(e',i'))|(x,(e,i)))=1\!-\!(1\!-\!r)\Big[1\!-\!\sum_{i'\neq i}\Big(1\!-\!\frac{N\!-\!2}{N\!-\!1}p_\epsilon(x_i,x_{i'},e)\Big)\prod_{j\neq i,i'}p_\epsilon(x_i,x_j,e)\Big] \\<1.
\end{split}
\label{eq:multicoll}
\end{equation}
A solution is to add the following lifting move,
$$
(1-r)\delta(e+e')\delta(i-i')\delta(x-x')\sum_{k=2}^{N-1}\sum_{\substack{1\leq j_1<\dots<j_k\leq N}}\prod_{m=1}^k(1-p_\epsilon(x_i,x_{j_m},e))\prod_{\substack{j\not\in (i,j_1,\dots,j_k)}}p_\epsilon(x_i,x_j,e),
$$
which replaces the \emph{rejected} physical move by a flip of
$e$. This however comes at the cost of extending the set $\mathcal{D}$
to include backward moves, i.e. $\{(-1,0),(0,-1)\}$. Fortunately, in
the infinitesimal limit $\epsilon\to0$ corresponding to the
continuous-time limit, this multiple rejection term is of order
$O(\epsilon^2)$, as
$ p_\epsilon(x_i,x_{j}, e)\stackrel{\epsilon\to 0}{\sim}
1-\epsilon\langle \nabla u(x_i,x_j), e\rangle_+$\footnote{We will note
  $\langle a,b\rangle_+\coloneqq \max(0,a\cdot b)$ and
  $\langle a,b\rangle_-\coloneqq -\min(0,a\cdot b)$.}, whereas the other
terms in the proposal distribution $q_\epsilon$ are at least of order
$O(\epsilon)$,
\begin{equation}
  \begin{aligned}
&q_\epsilon((x',(e',i')))|(x,(e,i)))\\
                  &=(1-r) \delta(e-e')\delta(i-i')\delta(x_i'+\epsilon e - x_i)\left(\prod_{j\neq i}\delta(x'_j-x_j)\right)\left(1-\epsilon\sum_{j\neq i}\langle \nabla u(x_i,x_j), e\rangle_+\right)\\
                  &+(1-r)\delta(e-e')(1-\delta(i-i'))\delta(x-x')\epsilon\langle\nabla u(x_i,x_{i'}, e\rangle_+
                  +r\delta(x-x')\mu_\setvar(\dd e',\dd i')+O(\epsilon^2)
  \end{aligned},
  \label{eq:inf_q_a}
\end{equation}
leading in this infinitesimal limit to
$$
\int_{\set\times\setvar} q_\epsilon((x',v')|(x,v))\dd x'\dd
v' = \int_{\set\times\setvar} q_\epsilon((x,v)|(x',v'))\dd
x'\dd v' = 1 - O(\epsilon^2).$$ Thus, Markov processes generated
by ECMC are composed of chains of ballistic trajectories
following the direction $e$ of spheres successively set by the label
$i$, updated at \emph{events} ruled by Poisson clocks stemming from
the continuous-time limit and of total rate
$\sum_{j\neq i}\langle\nabla u(x_i,x_j)\cdot e\rangle_+$. An illustration can
be found in Figure \ref{fig:evolution-pdmp}. As the acceptance
function $a$ is always returning $1$, these schemes have been referred
to as rejection-free. A pseudocode implementation is exhibited in
Algorithm \ref{alg:ECMC_poisson} and shows how to sample the sequence
of ballistic trajectories separated by the Poisson events.

      \begin{algorithm}
        \caption{ECMC implementation for soft disks outputting a set $S$ of $n$ samples}
        \label{alg:ECMC_poisson}
        \begin{algorithmic}          
          \State Set $S=\{\}$
          \State{Set $x \in \set$}
          \For{$k = 1$ to $k=n$}          
          \State Sample uniformly $(e,i) \in \{(1,0), (0,1)\} \times \llbracket 1, N \rrbracket$
          \State Set $E_R$ an exponential random variable with parameter $1$          
          \State Set $\Delta_R = -\frac{1}{r}\text{ln}E_R$       
          \While{$\Delta_R > 0$}
          \For{$j = 1$ to $j=N$, $j \neq i$}
          \State Set $E_j$ an exponential random variable with parameter $1$
          \State Compute $\Delta_{j}$ such that $\int_{0}^{\Delta_{j}} \langle \nabla u(x_i+se,x_j),e\rangle_+ \dd s = -\text{ln}(E_j)$  \EndFor
          \State Set $\Delta_{\text{Ev}},j_{\text{Ev}} = \min_{\substack{j \neq i}}(\Delta_j),\text{argmin}_{\substack{j \neq i}}(\Delta_j)$, $x_i \gets x_i + e\min(\Delta_R, \Delta_{\text{Ev}})$
          \If {$\Delta_R > \Delta_{\text{Ev}}$}
          \State Set $(e,i) \gets (e,j_{\text{Ev}})$          
          \State Set $\Delta_R \gets \Delta_R - \Delta_{\text{Ev}}$
          \EndIf
          \EndWhile
          \State Add $x$ to $S$
          \EndFor
          
          \Return S
        \end{algorithmic}
      \end{algorithm}

      As can be seen from the above informal derivation, the
      description of ECMC schemes in term of an infinitesimal limit of
      some finite schemes may appear cumbersome. In the next section,
      after introducing piecewise deterministic Markov processes
      (PDMP), we show how they offer a robust analytical framework for
      the analytical description of such scheme, making the derivation
      of some properties, e.g. invariance of a given distribution,
      more straightforward and allowing to properly describe schemes
      relying on a fixed-time refreshment.

\section{Piecewise deterministic Markov processes in ECMC algorithms for multiparticle systems}
\label{sec:pdmp}
\subsection{Definition of a PDMP}

Piecewise Deterministic Markov Processes have been formalized by Davis
in his seminal paper \cite{Davis_1984} and book
\cite{Davis_1993}. Briefly, and to fit to our setting, a PDMP
$\left\{ \pdmpX_t, \pdmpVar_t\right\}_{t \geq 0}$, defined on a space
$\set \times \setvar$, refers to a Markov process composed of
ballistic trajectories whose succession is ruled by a Poisson
process. In more details, once an initial state
$(X_0,V_0)\in \set \times \setvar$ is set, the process evolves
ballistically according to a deterministic differential flow
$(\phi_t)_{t \geq 0}$, until an event occurs or the process reaches
the domain boundary. Events are characterized by their rate
$\lambda: \set \times \setvar \rightarrow \mathbb{R}_+$ and a Markov
kernel $Q$, defined on
$\left( \set \times \setvar, \mathcal{B}(\set \times \setvar)\right)$,
which updates $\left\{ \pdmpX_t, \pdmpVar_t\right\}$. At the boundary,
for $(x,v)\in\partial\Omega\times\setvar$, the differential flow would
cause an exit from $\Omega\times\setvar$ for the subset
$\Gamma^* =
\{(z,v')\in\partial\Omega\times\setvar;\exists (t,(x,v))\in\mathbb{R}_+\times\set\times\setvar, \phi_t(x,v)=(z,v')\}$,
referred in the following as the exit boundary. On this exit boundary,
it is then another Markov kernel $Q^b$, called the boundary kernel and
defined on $\left(\Gamma^*, \mathcal{B}(\Omega\times\setvar)\right)$, which
updates the process.

For $(\pdmpx,\var) \in (\set \times \setvar)$ and $f$ in a suitable
functional space, say continuously differentiable function as an
example \footnote{The extended domain $\mathcal{D}(\mathcal{A}$) of
  the generator corresponds to functions $f$ such that
  $(f(X_t,V_t)-f(X_0,V_0)-\int_0^t\mathcal A f(X_s)\dd s)_{t\ge0})$ is
  a local martingale (\cite{Davis_1993}). The description of such
  domain $\mathcal D(\mathcal A) $, and a suitable core \cite[Th. 5.5,
  ]{Davis_1993}, proves to be a strenuous task for almost all considered
  PDMP-MCMC schemes (see \cite{Durmus_2018,Bierkens_2019}). Here we
  consider the case where we may associate to $(X_t,V_t)$ a strongly
  continuous semigroup so that $\mathcal A$ (associated to its domain
  $\mathcal D(\mathcal A) $) is the strong generator of our process.
  The fact that the continuous and continuously differentiable
  functions form a core follows from the same approximation procedure
  that in \cite[Prop. 21, Prop 23]{Durmus_2018} for the Bouncy
  Particle Sampler, for which one of the difficulty here may be that the jump
  rate may be non smooth.}, on $(\set \cup \partial\set)\times \setvar$,
the infinitesimal generator (or strong generator \cite{Davis_1993}).
associated with a PDMP is,
\begin{equation}
  \mathcal{A}f(\pdmpx,\var) = D_{\phi}f(\pdmpx,\var) +
  \lambda(\pdmpx,\var) \left(\int_{\set\times\setvar}  f(\pdmpx
    ',\var ')Q((\pdmpx,\var), (\mathrm{d}\pdmpx',\mathrm{d}\var')) -
    f(\pdmpx,\var) \right) 
\end{equation}
with $D_{\varphi}$ defined as,
\begin{equation}
  D_{\phi}f(\pdmpx,\var) =\left\{
    \begin{array}{ll}
   \underset{t\rightarrow 0^+}{\lim} \frac{f(\phi_t(\pdmpx,\var)) - f(\pdmpx,\var)}{t} \; &\mbox{ if this limit exists.} \\
 0 &\mbox{ otherwise.}
	\end{array}\right.
    \end{equation}
Now, on the exit boundary $(\pdmpx,v) \in \Gamma^*$, we have the boundary condition, 
    \begin{equation}
      f(\pdmpx,v) = \int_{\setvar}f(\pdmpx', v')Q^b((\pdmpx,v), (\dd x',\dd v')).
    \end{equation}
    Note that it is actually not necessary to specify $Q^b$ for
    boundary points which the process never actually hits and in the
    following we will note this set of reachable exit boundary points
    $\Gamma = \{(z,v')\in \Gamma^*; \exists (x,v)\in
    \Omega\times\setvar, P(\text{No event occured before reaching
      $(z,v')$ starting from $(x,v)$})>0 \}$, as in \cite{Davis_1984}.

      \subsection{Generator characterisation of PDMP in ECMC
        algorithms for multiparticle systems}

      We now characterize the stochastic processes produced by ECMC
      algorithms as PDMP, in the first and most common forms of ECMC
      schemes, as introduced in \cite{Bernard_2009,Michel_2014}. A
      PDMP characterisation can be done either algorithmically, as
      done up until now in statistical physics and as presented in
      section \ref{sec:ecmc_pres}, or by its generator as presented in
      the previous section and as done more recently in the context of
      Bayesian inference \cite{Bouchard_2018,Bierkens_2019}. In this
      section, we exhibit a generator characterisation of PDMP in
      ECMC, which is particularly helpful to prove the invariance of a
      given probability distribution.

      We first present a generator description common to previous PDMP
      characterisation of sampling algorithms and which is only
      valid for refreshment relying on a Poisson process. We then introduce
      another valid writing, which deals with the refreshment part as
      a boundary effect. In that way, it allows for more freedom in
      its choice, while better reflecting the algorithmic
      implementation based on fixed-time refreshment schemes as
      popularly used in statistical physics.

\subsubsection{Standard exponential refreshment strategy}
      
{\noindent \bf Differential flow $\phi$.} After extension of the state
space from $x\in\Omega\sim\pi$ to
$(x,v)\in\Omega\times\setvar\sim\pi\otimes\mu_\setvar$, the process
$(X_t,V_t)$ generated through ECMC is first characterized by the
following differential flow $\phi_t$ for all
$(x,v)\in\set\times\setvar$ and $t\geq 0$,
\begin{equation} \label{eq:deterministic_flow}
  \phi_t(x=(x_k)_{k\in\setind}, v=(e,i)) = ((x_1,\dots, x_{i-1},x_i + te,x_{i+1},\dots,x_N), v=(e,i)).
  \end{equation}
  This differential flow, translating the $i$-th sphere along $e$ is
  interrupted at events, ruled by the pairwise interactions and
  refreshment, and where only the lifting variable is updated through
  the Markov kernel $Q$. 

{\noindent \bf Markov kernel $Q$.}
For all $(x=(x_k)_{k\in\setind},v=(e,i))\in\set\times\setvar$ and
$A\in\mathcal{B}(\set\times\setvar)$,
\begin{equation} \label{eq:markov_kernel}
  Q((x,(e,i)), A) = \sum_{\substack{k=1\\k\neq i}}^N\frac{\lambda_k(x,(e,i))}{\lambda(x,(e,i))}\int_{\setvar}\1_A((x,(e', i')))Q_k((e,i),(\dd e',\dd i'))+ \frac{\lambda_r}{\lambda(x,(e,i))}\mu_\setvar(A),
\end{equation}
where, for all $k\in\setind$,$k\neq i$,
\begin{equation} 
  \lambda_k(x,(e,i)) = \langle \nabla_{x_i} u(x_i,x_k), e\rangle_+
\label{eq:rate}
\end{equation}
is the $(ik)$-pairwise event rate, $\lambda_r\in \mathbb{R}_+$ is a homogeneous
refreshment rate, making the total event rate now
$\lambda = \sum_{\substack{k=1\\k\neq i}}^N\lambda_k + \lambda_r$, and
$(Q_k)_{k\in\setind}$ are Markov kernels defined on
$\setvar\times\mathcal{B}(\setvar)$, so that, for all
$e\in\{(1,0),(0,1)\}$ and $(i, i')\in\setind^2$,
\begin{equation} 
  Q_k((e,i), (\dd e',\dd i')) = \delta(e- e')\delta(k-i')\dd e'\dd i',
  \label{eq:lift}
  \end{equation}
  thus coding for the $k$-th sphere being the one translated along
  $e$ after an $(ik)$-pairwise event occured.

  {\noindent \bf Boundary Markov kernel $Q^b$.}
 For sphere systems, the
  exit boundary is
  \begin{equation}\Gamma^* = \{(x,(e,i))\in\partial\Omega\times\setvar;\exists
    j\in\setind,(x_i,x_j)\in \Gamma^e_{\text{pair}}\},
    \label{eq:gamma_star}
    \end{equation}
    with,
    \begin{equation}
      \Gamma^e_{\text{pair}}=\{(x,x')\in
  \partial\Omega_{\text{pair}}; \langle \nabla_x d(x,x'), e \rangle \leq
  0 \}.
  \label{eq:gamma_pair}
\end{equation} and, given \eq{deterministic_flow}, \eq{markov_kernel},
  \eq{rate} and \eq{lift}, the corresponding set of reachable exit
  boundary points $\Gamma$ can be determined. 
For $(x,v)\in\Gamma$, the boundary Markov kernel $Q^b$
  is then of the following form, with
  $A\in\mathcal{B}(\set\times\setvar)$,
  \begin{equation}
  Q^b((x,(e,i)),A) = \sum_{\substack{k=1\\k\neq i}}^N \frac{\1_{\Gamma^e_{\text{pair}}}(x_i,x_k)}{\sum_{\substack{k=1\\k\neq i}}^N\1_{\Gamma^e_{\text{pair}}}(x_i,x_k)} \int_{\setvar}\1_A((x,(e', i')))Q_k((e,i),(\dd e', \dd i')),
  \label{eq:Qb}
\end{equation}
with the $\{Q_k\}_{k\in\setind}$ the Markov kernels defined on
$\left(\setvar, \mathcal{B}(\setvar)\right)$ as in \eq{lift}.  Note
that this choice of $Q^b$ can be applied in case of tangential
($\langle\nabla_{x_i}d(x_i,x_k),e\rangle=0$) or multiple collisions,
but we could have excluded these points from the definition of $Q^b$
as the set they form is small and does not impact the invariant
stationary distribution. Thus, for $(x,(e,i))$ not at a tangential
collision, we could also have proposed the following choice,
$$  \tilde{Q}^b((x,(e,i)),A) = \sum_{\substack{k=1\\k\neq i}}^N \frac{n_-(x_i,x_k,e)}{\sum_{\substack{k=1\\k\neq i}}^Nn_-(x_i,x_k,e)} \int_{\setvar}\1_A((x,(e', i')))Q_k((e,i),(\dd e', \dd i')),
$$
with $n_-(\cdot)$ the unnormalized negative pairwise normal component,
\begin{equation}
\begin{aligned}
  n_-(x_i,x_k,e) &= 1_{\Gamma^e_{\text{pair}}}(x_i,x_k)\langle -\nabla_{x_i} d(x_i,x_k),e \rangle\\
  &= \1_{\partial\Omega_{\text{pair}}}(x_i,x_k)\langle \nabla_{x_i} d(x_i,x_k),e \rangle_-,  
\end{aligned}
\label{eq:normal_pair}
\end{equation}
which differs only from \eq{Qb} in a multiple collision situation,
where it is handled as when dealing with $n-$body interactions as done
in \cite{Harland_2017}. We expect these different choices to have an
impact while studying out-of-equilibrium processes though, but such
consideration is delayed for future work. We also refer the reader to
the recent work of \cite{Chevallier2021} for a more general
consideration into PDMP samplers for piecewise continuous densities.

Eventually, it
leads to the following condition on the boundary, for
$(x,(e,i))\in\Gamma$ and $f$ a continuous and continuously
differentiable function on  $(\set \cup \partial\set)\times \setvar$,
  \begin{equation}
    f(\pdmpx,(e,i)) = \sum_{\substack{k=1\\k\neq i}}^Nf(\pdmpx, (e,k))\frac{\1_{\Gamma^e_{\text{pair}}}(x_i,x_k)}{\sum_{\substack{k=1\\k\neq i}}^N\1_{\Gamma^e_{\text{pair}}}(x_i,x_k)}.
    \label{eq:Hbound}
   \end{equation}

  {\noindent \bf Soft-sphere systems.} All in all, the following infinitesimal generator associated with
   PDMP generated by ECMC for soft-sphere systems comes down to, 
  with $f$ a continuous and continuously differentiable function on
  $(\set_S\cup\partial\set_S)\times\setvar$ and $(x,(e,i))\in\set_S\times\setvar$,  
  \begin{equation}
    \begin{aligned}
  \mathcal{A}_Sf(\pdmpx,(e,i)) = \langle \nabla_{x_i} f(x,(e,i)), e\rangle 
  &+\sum_{k=1}^N \langle \nabla_{x_i} u_S(x_i,x_k), e\rangle_+ \left\{f(\pdmpx, (e,k))-f(\pdmpx,(e,i))\right\}\\
  &+ \lambda_r\left(\int_\setvar f(x,(e', i'))\dd\mu_{\setvar}((e', i')) - f(\pdmpx, (e,i))\right).
\end{aligned}
\label{eq:Asoft}
\end{equation}
For soft-sphere systems, $\Gamma_S$ is empty, as the event rate
diverges as a pair distance goes to $d_{\text{pair}}=0$ \eq{rate}, and
the boundary Markov kernel is without any object.

{\noindent \bf Hard-sphere systems.} For hard-sphere systems, the
infinitesimal generator associated, with $f$ a continuous and
continuously differentiable function on
$(\set_H\cup\partial\set_H)\times\setvar$ and
$(x,(e,i))\in\set_H\times\setvar$,
\begin{equation} \label{eq:Ahard}
  \mathcal{A}_{H}f(\pdmpx,(e,i)) = \langle \nabla_{x_i} f(x,(e,i)), e\rangle 
  + \lambda_r\left(\int_\setvar f(x,(e', i'))\dd\mu_{\setvar}((e', i')) - f(\pdmpx, (e,i))\right).
\end{equation}   
For hard-sphere systems, $\Gamma_H = \Gamma^*_H$, as the events can
only be triggered by refreshments and they do not impact a point
reachability. Then, for $(x,v)\in\Gamma_H$, the boundary Markov kernel
is the one defined in \eq{Qb} and eventually leading to the condition
\eq{Hbound} on the reachable exit boundary $\Gamma_H$. Thus the
hardcore repulsions are only appearing as boundary effects.

\subsubsection{Boundary refreshment strategy}

The PDMP description where the refreshment part is treated as part of
the jump process is the most common one. It can however not be used to
study ECMC implementations where the refreshment process is not an
exponential process. For instance, the fixed-time refreshment is a
common and useful practice in statistical physics, all the more while
dealing with periodic boundaries. It can also alleviate some
difficulties in the computation of the event times set by the pairwise
rates $(\lambda_k)_{k\in\setind}$.

Therefore we now explain how to treat the refreshment as a boundary
effect by adding an additional variable $l\in\mathcal{L}$, with
$\mathcal{L} = ]0,+\infty[$, $\partial \mathcal{L} = \{0\}$,
associated with the measure $\mu_{L}$, extended by continuity to
$\tilde{\mu}_L$ on $\mathcal{L}\cup\partial\mathcal{L}$. The state
space $\set$ is now extended to $\set\times\mathcal{L}\times\setvar$
and the process $(X_t,L_t,V_t)$ now targets as a stationary
distribution $\pi\times\mu_L\times\mu_\setvar$. We then discuss a
broader range of possible refreshment strategies.

   {\noindent \bf Differential flow.} The differential flow $\varphi_t$ for
   all $(x,l,(e,i))\in\set\times\mathcal{L}\times\setvar$ and
   $t\geq 0$ is now
   \begin{equation}
     \varphi_t(x,l,(e,i)) = ((x_1,\dots,x_i+te,\dots,x_N),l-t, (e,i)).
     \end{equation}
     {\noindent \bf Markov kernel $Q$.} The Markov kernel $Q$ is, for all
     $(x,l,(e,i))\in\set\times\mathcal{L}\times\setvar$ and
     $A\in\mathcal{B}(\set\times\mathcal{L}\times\setvar)$,
\begin{equation}
Q((x,l,(e,i)), A) = \sum_{\substack{k=1\\k\neq i}}^N\frac{\lambda_k(x,(e,i))}{\lambda(x,(e,i))}\int_{\setvar}\1_A((x,l,(e',i')))Q_k((e,i),(\dd e', \dd i')),
\end{equation}
where $\{\lambda_k(x,(e,i))\}_{k\neq i}$ are the pairwise rates
defined in \eq{rate}, the total event rate being now
$\lambda = \sum_{k=1;k\neq i}^N\lambda_k$ and $(Q_k)_{k\in\setind}$ are the
Markov kernels defined on $\setvar\times\mathcal{B}(\setvar)$ in
\eq{lift}.

{\noindent \bf Boundary kernel $Q^b$.}
We first define the set of exit boundary points,
\[\Gamma^* = \{(x,l,(e,i))\in\partial(\Omega\times\mathcal{L})\times\setvar;\exists j\in\setind,(x_i,x_j)\in \Gamma^e_{\text{pair}} \text{ or } l=0\}
  \]
  and $\Gamma^e_{\text{pair}}$ is defined as in \eq{gamma_pair}.
  The corresponding set $\Gamma$ of reachable exit
  boundary points is included in $\Gamma^*$, and, for
  $(x,l,v)\in \Gamma\times\setvar$, the boundary Markov kernel $Q^b$
  is, with $A\in\mathcal{B}(\set\times\mathcal{L}\times\setvar)$,
  \begin{equation}
    \begin{aligned}
      & Q^b((x,l,(e,i)),A) = 
      \1_{\partial\mathcal{L}}(l)\int_{\mathcal{L}}\ 1_A((x,l',(e',i')))R(l,\dd l')\dd\mu_\setvar((e', i'))\\      
  &+   (1-\1_{\partial\mathcal{L}}(l)) \sum_{\substack{k=1\\k\neq i}}^N \frac{\1_{\Gamma^e_{\text{pair}}}(x_i,x_k)}{\sum_{\substack{k=1\\k\neq i}}^N\1_{\Gamma^e_{\text{pair}}}(x_i,x_k)} \int_{\setvar}\1_A((x,l,(e',i')))Q_k((e,i),\dd(e',i'))
 \end{aligned}
\end{equation}
with $R$ a Markov kernel defined on
$\mathcal{L}\times\mathcal{B}(\mathcal{L})$ and the
$\{Q_k\}_{k\in\setind}$ the Markov kernels defined on
$\left(\setvar, \mathcal{B}(\setvar)\right)$ as in \eq{lift}. One
could naturally consider more general choice for $R$ than just a
kernel acting on the refreshment time $l$. Also, as previously
mentioned, we could have given an explicit definition of the boundary
kernel on only points forming a non-small set, excluding tangential,
multiple collisions or coincidental collision and refreshment.

Eventually, the infinitesimal generator comes down to, with $f$ a
continuous and continuously differentiable function on
$((\set\times\mathcal{L})\cup\partial(\set\times\mathcal{L}))\times\setvar$ and
$(x,l,(e,i))\in\set\times\mathcal L\times\setvar$,
  \begin{equation}
    \begin{aligned}
  \mathcal{A}f(\pdmpx,l,(e,i)) =& \langle \nabla_{x_i} f(x,l,(e,i)), e\rangle -\partial_lf(x,l,(e,i))
  \\&+\sum_{k=1}^N \lambda_k(\pdmpx,l,(e,i)) \left\{f(\pdmpx, l,(e,k))-f(\pdmpx,l,(e,i))\right\},
\end{aligned}
\label{eq:AsoftB}
\end{equation}

And, we have the condition on the boundary for
$(x,l,(e,i))\in \Gamma\times\setvar$ and
for $f$ a continuous and continuously differentiable function on
$((\set\times\mathcal{L})\cup\partial(\set\times\mathcal{L}))\times\setvar$,
\begin{equation}
  \begin{aligned} \label{eq:fHB}
    f(x,l,(e,i)) =  &\1_{\partial \mathcal{L}}(l) \int_{\mathcal{L}} f(x,l',(e',i'))R(l,\dd l')\dd\mu_\setvar((e',i'))\\
&+(1-\1_{\partial \mathcal{L}}(l))\sum_{\substack{k=1\\k\neq i}}^N\frac{\1_{\Gamma^e_{\text{pair}}}(x_i,x_k)}{\sum_{\substack{k=1\\k\neq i}}^N\1_{\Gamma^e_{\text{pair}}}(x_i,x_k)} f(x,l,(e,k)).
    \end{aligned}
  \end{equation}
  This shift in description from an exponential jump process to a
  boundary effect could also be carried on regarding the events
  stemming from the pairwise interactions and ruled by the rates
  $(\lambda_k)_{k\in\setind}$ \eq{rate}. It nicely reflects the
  picture of an energy reservoir emptied along the differential flow
  from the positive energy increment, as described in the first works
  introducing these algorithmic methods
  \cite{Michel_2014,Michel_2015}. We will present conditions on $R$ to ensure the correct invariance towards our target measure $\pi \times \mu_{L} \times \mu_{\setvar}$ in the next section.

\section{Invariance of the equilibrium distribution}
\label{sec:inv}

The generator is an efficient tool to prove invariance of a measure
w.r.t. a given process (e.g. \cite[Prop. 34.7]{Davis_1993}). It will
be normally required to do the formal effort to characterize the core
of its generator (e.g. \cite[Cor. 22]{Durmus_2018}). As it is not the
heart of our problematic, we will not detail the approximation
procedure, as described via \cite[Prop. 23, Cor.24]{Durmus_2018} or
\cite{H2021}, so that $\pi\otimes\mu_{\setvar}$ is shown to be left
invariant by the PDMP by means of its infinitesimal generator
\eqref{eq:Asoft} applied to continuously differentiable
functions. Note that as we consider the torus, $f$ is also bounded so
that we have no problem defining the condition on the boundary and
thus requires no additional condition on $Q_b$.

\subsection{Standard exponential refreshment strategy}

This comes down to show, with $f$ a continuous and continuously
differentiable function on $(\set\cup\partial\set)\times\setvar$, that,
\begin{equation}
  \int_{\set \times \setvar}
    \mathcal{A}f(x,(e,i))\pi(x)\mathrm{d}x
    \dd\mu_\setvar(e,i) = 0
\end{equation}
Remark here that, from a dynamical point of view, starting from a
measure with a nonzero density with regards to $\pi$ is important, so
as not to charge overlapping configurations in the soft-sphere case
or spheres in contact in the hard-sphere one.

Expliciting $\mathcal{A}$ from \eq{Asoft} for soft spheres or
\eq{Ahard} for hard ones, we obtain
  \begin{equation}
    \begin{aligned}
      \int_{\set\times\setvar}\mathcal{A}f(\pdmpx,(e,i))\dd\pi(x)\dd\mu_\setvar(e,i) &=
       \int_{\set\times\setvar}\dd\pi(x)\dd\mu_\setvar(e,i)\langle \nabla_{x_i} f(x,(e,i)), e\rangle \\
  &+\int_{\set\times\setvar}\dd\pi(x)\dd\mu_\setvar(e,i)\sum_{\substack{k=1\\k\neq i}}^N \langle e, \nabla_{x_i} u(x_i,x_k)\rangle_+ \left\{f(\pdmpx, (e,k))-f(\pdmpx,(e,i))\right\}\\
  &+ \lambda_r\int_{\set}\dd\pi(x)\left(\int_\setvar\dd\mu_{\setvar}(e',i') f(x,(e',i')) - \int_{\setvar}\dd\mu_\setvar(e,i)f(\pdmpx, (e,i))\right)\\
\end{aligned}
\end{equation}
The refreshment term cancels itself. Now, by integration by parts,
\begin{equation}
  \begin{aligned}
        &= \int_{\set\times\setvar}\dd x\dd\mu_\setvar(e,i) \nabla_{x_i}\cdot(f(x,(e,i))\pi(x)e)
-\int_{\set\times\setvar}\dd x\dd\mu_\setvar(e,i) f(x,(e,i))\langle \nabla_{x_i} \pi(x), e\rangle
  \\
  &+\int_{\set\times\mathcal{D}}\dd x\dd\mu_\mathcal{D}(e) \frac 1N\sum_{i=1}^N\sum_{\substack{k=1\\k\neq i}}^N\pi(x) \langle e, \nabla_{x_i} u(x_i,x_k)\rangle_+ \left\{f(\pdmpx, (e,k))-f(\pdmpx,(e,i))\right\}.
\end{aligned}
\label{eq:int_part}
\end{equation}
Key point of these schemes, we make use of the pairwise mirror
symmetry $\nabla_{x_i}u(x_i,x_k) = -\nabla_{x_k}u(x_i,x_k)$ to show
the compensation of the transport by the events,
\begin{equation}
  \begin{aligned}
  &= \int_{\set\times\setvar}\dd x\dd\mu_\setvar(e,i) \nabla_{x_i}\cdot(f(x,(e,i))\pi(x)e)-\int_{\set\times\setvar}\dd x\dd\mu_\setvar(e,i) f(x,(e,i))\langle \nabla_{x_i} \pi(x), e\rangle
  \\
   &+\int_{\set\times\mathcal{D}}\dd x\dd\mu_\mathcal{D}(e) \frac{1}{2N}\sum_{i=1}^N\sum_{\substack{k=1\\k\neq i}}^N\pi(x) \langle  \nabla_{x_i} u(x_i,x_k), e\rangle \left\{f(\pdmpx, (e,k))-f(\pdmpx,(e,i))\right\}\\
  &= \int_{\set\times\setvar}\dd x\dd\mu_\setvar(e,i) \nabla_{x_i}\cdot(f(x,(e,i))\pi(x)e)-\int_{\set\times\setvar}\dd x\dd\mu_\setvar(e,i) f(x,(e,i))\langle \nabla_{x_i} \pi(x), e\rangle
  \\
  &+\int_{\set\times\mathcal{D}}\dd x\dd\mu_\mathcal{D}(e) \frac{1}{2N}\left(\sum_{i=1}^N \langle  \nabla_{x_i}\pi(x), e\rangle f(\pdmpx, (e,i))+\sum_{k=1}^N\langle  \nabla_{x_k}\pi(x), e\rangle f(\pdmpx,(e,k))\right)\\
    &= \int_{\set\times\setvar}\dd x\dd\mu_\setvar(e,i) \nabla_{x_i}\cdot(f(x,(e,i))\pi(x)e).
  \end{aligned}
\label{eq:compens}
\end{equation}
By the divergence theorem, the first term encodes for the effects on
the boundary $\partial \Omega\times \setvar$,
\begin{equation}
  \begin{aligned}
  \int_{\set\times\setvar}\dd x\dd\mu_\setvar(e,i) \nabla_{x_i}\cdot(f(x,(e,i))\pi(x)e)=&
\int_{(\partial\Omega\times\setvar)\setminus\Gamma^*}\dd x\dd\mu_{\setvar}(e,i) f(x,(e,i))\tilde{\pi}(x)
 \left\langle n_{i}(x,(e,i)), e\right\rangle.\\
&+\int_{\Gamma}\dd x\dd\mu_{\setvar}(e,i) f(x,(e,i))\tilde{\pi}(x)
 \left\langle n_{i}(x,(e,i)), e\right\rangle\\
&+\int_{\Gamma^*\setminus\Gamma}\dd x\dd\mu_{\setvar}(e,i) f(x,(e,i))\tilde{\pi}(x)
\left\langle n_{i}(x,(e,i)), e\right\rangle,
\end{aligned}
 \label{eq:divbound}
\end{equation}
with $n_i$ the $i$-th component of the local outward normal, 
$$n_i(x,(e,i)) =-\sum^N_{\substack{k=1\\k\neq i}}\1_{\Omega_{\text{pair}}}(x_i,x_k)\nabla_{x_i}d(x_i,x_k)$$
which can be rewritten as,
\begin{equation}
  n_i(x,(e,i)) = \sum^N_{\substack{k=1\\k\neq i}} (\1_{\Gamma^e_{\text{pair}}}(x_k,x_i)\nabla_{x_k}d(x_i,x_k)-\1_{\Gamma^e_{\text{pair}}}(x_i,x_k)\nabla_{x_i}d(x_i,x_k)).
  \label{eq:normal_rel}
\end{equation}

For soft-sphere systems, $\tilde{\pi}(x)$ is $0$ for
$x\in \partial \Omega$ \eq{pisoft}, yielding \eq{divbound} to sum up
to $0$ and the invariance of $\pi_S\times\mu_V$. The situation is
different for hard-sphere ones where $\Gamma_H =
\Gamma^*_H$. Using the relation on the boundaries \eq{Hbound}, the relation
\eq{normal_rel} and the definition of $\Gamma_H$, we get
\begin{equation}
  \begin{aligned}
=&
\int_{(\partial\Omega\times\setvar)\setminus\Gamma_H}\dd x\dd\mu_{\setvar}(e,i) f(x,(e,i))
\sum^N_{\substack{k=1\\k\neq i}}\1_{\Gamma^e_{\text{pair}}}(x_k,x_i)\langle\nabla_{x_k}d(x_i,x_k),e\rangle\\
&+ \int_{\Gamma_H}\dd x\dd\mu_{\setvar}(e,i) f(x,(e,i))
 \sum^N_{\substack{k=1\\k\neq i}}\1_{\Gamma^e_{\text{pair}}}(x_k,x_i)\langle \nabla_{x_k}d(x_i,x_k),e\rangle\\
&- \int_{\Gamma_H}\dd x\dd\mu_{\setvar}(e,i) \sum_{\substack{j=1\\j\neq i}}^Nf(x,(e,j))\frac{\1_{\Gamma^e_{\text{pair}}}(x_i,x_j)}{\sum_{\substack{j=1\\j\neq i}}^N\1_{\Gamma^e_{\text{pair}}}(x_i,x_j)}
 \sum^N_{\substack{k=1\\k\neq i}}\1_{\Gamma^e_{\text{pair}}}(x_i,x_k)\langle\nabla_{x_i}d(x_i,x_k),e\rangle.
\end{aligned}
\end{equation}
Merging the first two terms and simplyfing the third as multicollisions form a small set,
\begin{equation}
  \begin{aligned}
=&
\int_{\partial\Omega\times\setvar}\dd x\dd\mu_{\setvar}(e,i) f(x,(e,i))
\sum^N_{\substack{k=1\\k\neq i}}\1_{\Gamma^e_{\text{pair}}}(x_k,x_i)\langle\nabla_{x_k}d(x_i,x_k),e\rangle\\
&- \int_{\Gamma_H}\dd x\dd\mu_{\setvar}(e,i) \sum_{\substack{j=1\\j\neq i}}^Nf(x,(e,j))\1_{\Gamma^e_{\text{pair}}(x_i,x_j)}\langle \nabla_{x_i} d(x_i,x_j) ,e\rangle.
\end{aligned}
\end{equation}
As $\Gamma_H \subset \partial \Omega\times\setvar$,
\begin{equation}
  \begin{aligned}
=&
\int_{\partial\Omega\times\setvar}\dd x\dd\mu_{\setvar}(e,i) f(x,(e,i))
\sum^N_{\substack{k=1\\k\neq i}}\1_{\Gamma^e_{\text{pair}}}(x_k,x_i)\langle\nabla_{x_k}d(x_i,x_k),e\rangle\\
&- \int_{\partial\Omega\times\setvar}\dd x\dd\mu_{\mathcal{D}}(e)\frac 1N \sum_{i=1}^N \sum_{\substack{j=1\\j\neq i}}^N\1_{\Gamma_H}(x,(e,i))f(x,(e,j))\1_{\Gamma^e_{\text{pair}}(x_i,x_j)}\langle \nabla_{x_i} d(x_i,x_j) ,e\rangle,
\end{aligned}
\end{equation}
which identifies with,
\begin{equation}
  \begin{aligned}
=&
\int_{\partial\Omega\times\setvar}\dd x\dd\mu_{\setvar}(e,i) f(x,(e,i))
\sum^N_{\substack{k=1\\k\neq i}}\1_{\Gamma^e_{\text{pair}}}(x_k,x_i)\langle\nabla_{x_k}d(x_i,x_k),e\rangle\\
&- \int_{\partial\Omega\times\setvar}\dd x\dd\mu_{\setvar}(e,j) f(x,(e,j))\sum^N_{\substack{i=1\\i\neq j}}\1_{\Gamma^e_{\text{pair}}(x_i,x_j)}\langle \nabla_{x_i} d(x_i,x_j) ,e\rangle\\
=&0,
\end{aligned}
\end{equation}
leading to the invariance of $\pi_H\times\mu_\setvar$ in the hard-sphere
case. The invariance can be obtained in the same manner when
using $\tilde{Q}^b$ by noting the relation,
\begin{equation}
  \langle n_i(x,(e,i)), e\rangle =\sum^N_{\substack{k=1\\k\neq i}} (n_-(x_i,x_k,e)-n_-(x_k,x_i,e)).
  \label{eq:normal_rel_n_minus}
\end{equation}

\subsection{Refreshment as a boundary effect}

We show the invariance of the target distribution
$\pi \times \mu_{L} \times \mu_{\setvar}$ by verifying the invariance
condition on \eqref{eq:AsoftB}. With $f$ a continuous and continuously
differentiable function on
$((\set\times\mathcal{L})\cup\partial(\set\times\mathcal{L}))\times\setvar$ and
$(x,l,(e,i))\in\set\times\mathcal L\times\setvar$, we get by
integration by part and running similar computations as done in \eq{int_part}
and \eq{compens},

  \begin{equation}
    \begin{aligned}
      \int_{\set\times\mathcal{L}\times\setvar}\mathcal{A}f(\pdmpx,l,(e,i))\dd\pi(x)\dd \mu_L(l)\dd\mu_\setvar(e,i)
      &=
      \int_{\set\times\mathcal{L}\times\setvar}\dd x\dd \mu_L(l)\dd\mu_\setvar(e,i) \nabla_{x_i}\cdot(f(x,l,(e,i))\pi(x)e)\\
      &- \int_{\set\times\mathcal{L}\times\setvar}\dd \pi(x)\dd \mu_L(l)\dd\mu_\setvar(e,i) \partial_{l} f(x,l,(e,i))).   
    \end{aligned}
    \label{eq:cond_refB}
\end{equation}
And, by the divergence theorem and integration by parts, we obtain,
\begin{equation}
  \begin{aligned}
    &= \int_{\partial{\set}\times\mathcal{L}\times\setvar}\dd x\dd \mu_L(l)\dd\mu_\setvar(e,i) f(x,l,(e,i))\tilde{\pi}(x)\langle n_i(x,(e,i)), e\rangle\\
    &+\int_{\set\times\setvar}\dd\pi(x) \dd\mu_\setvar(e,i) \tilde{\mu}_L(0)f(x,0,(e,i))
 -  \int_{\set\times\mathcal{L}\times\setvar}\dd\pi(x)\dd l \dd\mu_\setvar(e,i)(-\partial_l \mu_L(l)) f(x,l,(e,i))
  \end{aligned}
    \label{eq:condbound_refB}
\end{equation}
  Using condition \eq{fHB}, we obtain the following general
condition on the boundary refreshment kernel $R$ in order to set
\eq{condbound_refB} to 0,
\begin{equation}
 \int_{\set\times\mathcal{L}\times\setvar} \dd\pi(x) \dd\mu_\setvar(e,i) \tilde{\mu}_L(0) f(x, l, (e,i))R(0,dl)
=  \int_{\set\times\mathcal{L}\times\setvar}\dd\pi(x)\dd l \dd\mu_\setvar(e,i)(-\partial_l \mu_L(l)) f(x,l,(e,i)),
\end{equation}
Note that here the kernel $R$ has to compensate the transport term and
strongly depends on $\mu_L$, contrary to the boundary kernel $Q^b$
which can only depend on local information as the local
normal. This condition simplifies to, for $l\in\mathcal{L}$,
\begin{equation} \label{eq:condition_ref_kernel}
\tilde{\mu}_L(0) R(0,dl)
=  (-\partial_l \mu_L(l)) \dd l.
\end{equation}
It is equivalent to requiring $\mu_L$ to be of the form,
  \begin{equation}
    \mu_L(l) = h(l) \1_{D}.
  \end{equation}
  where $D \subset{\mathcal{L}}$ and $h$ a decreasing function on $D$
  so that $\lim_{l\to 0^+}h(l)>0$ and $\int_{\mathcal{L}}\frac{-\partial_l \mu_L(l)}{\tilde{\mu}_L(0) } \dd l=1$.
  
  Naturally we directly recover processes currently used in most
  algorithms, i.e. the fixed-time refreshment or the exponential one,
  giving, for $l \in \mathcal{L}$,
$$\left\{\begin{array}{llll}
           \text{fixed-time $T$: }& \mu_L (l)=\frac1T\1_{]0,T[} &\text{and } R(l,\dd l')= \delta(l' - T)\dd l',& \text{ with } T>0, D=]0,T] \\
           \text{exponential of rate $\lambda_r$: }& \mu_L (l)= \lambda_re^{-\lambda_r l}\1_{l>0} &\text{and } R(l,\dd l') = \lambda_re^{-\lambda_r l'}\1_{l'>0}\dd l',&  \text{ with } \lambda_r>0, D=\mathcal{L}
           \end{array}\right.
.$$
But also, building on the flexibility of the boundary description, new
refreshment strategies are possible
  \begin{itemize}
  \item $\mu_L(l) = A(T-l)^k \1_{0< l \leq T}$, and refreshment 
    $R(0,\dd l)=\frac {k}{T^k}(T-l)^{k-1} \1_{0< l \leq T}$,
  \item $\mu_L(l) = \frac{A}{(T+l)^k} \1_{0<l}$, and refreshment
    $R(0,\dd l)= \frac{kT^k}{(T+l)^{k+1}} \1_{0<l}$,
  \item$\mu_L(l) = Ae^{-Tl^k} \1_{0<l}$, and refreshment
    $R(0,\dd l)= kTl^{k-1}e^{-Tl^k} \1_{0<l}$,
  \item ...
  \end{itemize}
where $(T,k)\in \mathbb{R}_+^{*2}$, $l \in \mathcal{L}$ and $A$ is a suitable normalization constant.

We have thus provided a general framework, setting refreshment as
boundary type conditions, in order to allow new algorithms with
different refreshment strategies. It would be interesting to look at
the effect on the speed of convergence towards the equilibrium, or
equilibration of different observables of these algorithms. Remark
that we may also combine the usual Poisson refreshment with other
refreshment conditions coming from the boundary (at the expense of
adding other variables) to enrich refreshment strategies.

\section{Ergodicity of ECMC}
\label{sec:ergo}

Contrary to the random-walk Metropolis-Hasting algorithm addressed in
\cite{Diaconis_2011}, which generates a reversible Markov chain where
each sphere is allowed to move uniformly in a neighborhood of its
current position (if no collision occurs), the continuity of the
underlying topological state space imposes to take care of the
existence, not simply of a single path, but of a density of paths
connecting any two states. We can gain such a density by the
randomness of the jump times, and it can be achieved using events or
refreshments. Out of simplicity, we will consider here only
refreshments. It is a difficult and technical task to prove ergodicity
for this very ballistic process. We refer to the recent
\cite{Bierkens_2019,Durmus_2020} for such a study for the Zig-Zag
process and the Bouncy Particle Sampler.

It would be of course very interesting to get explicit exponential
speed of convergence towards equilibrium, suitably scaled with respect
to the number of spheres. However, our estimates are definitely too
crude to provide such an evaluation and we thus stay at a qualitative
level. We also provide an alternative reachability strategy, opening
new perspectives into getting coupling or uniform ergodicity
results. Note that, even for (kinetic) Langevin process in the
soft-sphere case, the only results at our disposal are based on
Lyapunov-type techniques \cite{Lu_2018}, only asserting exponential
convergence but no rates. We refer to
\cite{Bernard_2009,Bernard_2011,Michel_2014} for numerical evidence of
the efficiency of the ECMC for soft- and hard-sphere cases.

\subsection{Results and schemes of proofs}
\label{sec:ergo_ss}

We focus here on the usual case where the refreshment is not seen as a
boundary effect. Following \cite[Th. 6.1]{Meyn_1993}, and a recent
application in a similar context \cite{Bierkens_2019}, we show that
the PDMP is positive Harris recurrent and that some skeleton chain is
irreducible through the control of distances and probability
minorization to obtain a density of path connecting the initial and
final states. To this end, we define different paths depending on the
starting and final configurations, e.g. if the spheres are well
separated or not, which impacts how easy it is to define a connecting
path which moves each sphere sequentially. In the case of spheres
being not separated enough, we thus adapt the strategy of
\cite{Diaconis_2011} for proving the connectivity of their reversible
algorithm, while improving to some extent the density condition. We
then use the nice tools of \cite{Benaim_2015} to gain density of paths
from connectivity.

More precisely, the process $(X_t, V_t)_{t\geq 0}$ considered here is
non-evanescent due to the periodic boundary conditions of $\set_S$. It
is Harris recurrent if it is in addition a $\phi$-irreducible
T-process (\cite{Meyn_1993}, Theorem 3.2) and the positivity comes
from the existence of an invariant probability distribution
(\cite{Azema_1967,Getoor_1980}). Finally, a positive Harris recurrent
process with an invariant probability distribution is ergodic if some
skeleton chain is irreducible (\cite{Meyn_1993}, Theorem 6.1). This
leads to
\begin{theorem}
   If the density condition
   \begin{equation}\label{eq:density2}
  \exists \epsilon >0,  3N \leq \left\lfloor \frac{L}{2d_{\text{pair}}+\epsilon}\right\rfloor
    \left\lfloor \frac{L}{d_{\text{pair}}+\epsilon}\frac{\sqrt{3}}{3}\right\rfloor
 \end{equation}
 is satisfied, the PDMP $(X_t, V_t)_{t \geq 0}$ with the differential
 flow \eqref{eq:deterministic_flow}, the Markov kernel
 \eqref{eq:markov_kernel}, the event rates \eqref{eq:rate} and
 described by the generator \eq{Asoft} or \eq{Ahard}, is ergodic.
\end{theorem}
The density condition \eq{density2} simply ensures the possibility to
pack without contact $3N$ spheres of radius $d_{\text{pair}}$ in the
considered torus, via a hexagonal packing.

The distribution $\pi \otimes \mu_{\setvar}$ is invariant for the
PDMP. Following \cite{Benaim_2015}, to prove the irreducibility of the
process, we define the set of trajectories composed of $m\in\mathbb{N}^*$ jumps,
\begin{equation} \label{eq:index_set}
\begin{split}
\mathbb{T}_m = \left\{(\mathbf{t},\mathbf{v}); \; \mathbf{t} = \left(t_1,\dots, t_m \right) \in \mathbb{R}_+^m, \, \mathbf{v} = \left((e_0,i_0),(e_1,i_1), \dots, (e_m,i_m) \right) \in \setvar^{m+1}\right\},
\end{split}
\end{equation}
and the composite flow
$\phi_{\mathbf{t}}^{\mathbf{v}} = \phi_{t_m}^{(e_{m-1},i_{m-1})} \circ \dots
\circ\phi_{t_1}^{(e_0,i_0)}$ for
$(\mathbf{t},\mathbf{v}) \in \mathbb{R}_+^m\times\setvar^{m'}, m'\geq m$, with
$\phi_t^{(e,i)}:\Omega \to \Omega$ defined so that
$\phi_t(x,(e,i))=(\phi_t^{(e,i)}(x), (e,i))$. Our crucial tool will be

\begin{lemma}\label{lem:submersion}
  If there exists $\epsilon>0$,
  $ 3N \leq \left\lfloor \frac{L}{2d_{\text{pair}}+\epsilon}\right\rfloor
  \left\lfloor
    \frac{L}{d_{\text{pair}}+\epsilon}\frac{\sqrt{3}}{3}\right\rfloor$, for any
  pair
  $((x^{(0)},v^{(0)}),(x^{(f)},v^{(f)})) \in (\set\times\setvar)^2$,
  there exists a trajectory
  $(\mathbf{t}, \mathbf{v}) \in \mathbb{T}_{m}, m\in\mathbb{N}^*$,
  such that $v_0=v^{(0)}$, $v_m=v^{(f)}$ and
  $\phi_{\mathbf{t}}^{\mathbf{v}}(x_0) = x_f$ and the application
  $\displaystyle \bm{\tau} = (\tau_{k})_{k=1}^{m} \to \phi_{(t' -
    \sum_{k=1}^{m}\tau_{k})}^{v_m}\left(\phi_{\bm{\tau}}^{\mathbf{v}}(x^{(0)})\right)$
  is a submersion at $\mathbf{t}$ for some
  $t'>\sum_{i=1}^m t_i$.
\end{lemma}

Proving Lemma \ref{lem:submersion} comes down to designing a
trajectory $(\mathbf{t}, \mathbf{v})\in\mathbb{T}_{m},  m\in\mathbb{N}^*,$ connecting
$(x^{(0)},v^{(0)})$ and $(x^{(f)},v^{(f)})$ in which all the possible pairs
$v=(e,i) \in \setvar$ appear at least once,
i.e. $\setvar\subset\{v_k\}_{k=0}^{m}$. Furthermore, the resulting path on $\set$
can be deformed by using another sequence of times
$\bm{\tau} \in \mathbb{R}_{+}^{m}$ while reaching the same endpoint
$x_f$ given some additional time
$t'\in\mathbb{R}_+^*$.

Lemma \ref{lem:submersion} then ensures that the process can reach a
neighborhood of the endpoint $x_f$. As a consequence, the theorem 4.2
of \cite{Benaim_2015} applies. One may remark that they impose bounded jump rates, but it is only for the construction of their process, and it does not intervene in their proof of Theorem 4.2. Thus for all $t' > \sum_{i=1}^{m}t_i$, there
exist neighborhoods ${\cal X}_0$ of $x_0$, ${\cal X}_{f}$ of $x_f,$
and constants $c,\epsilon > 0$ such that
\begin{equation} \label{eq:abscontinuity}
\forall x \in {\cal X}_0, \, \forall (v,v') \in \setvar^2, \,  \forall t \in [t', t' + \epsilon], \; \mathbb{P}_{(x, v)}\left((X_t,V_t) \in \cdot \times \{v'\}\right) \geq c \mathrm{Leb}(\cdot \cap {\cal X}_f),
\end{equation}
Such lower bound has the following consequences (\cite[Lem 8,
Th. 5]{Bierkens_2019}): there exist a locally finite family of open
sets $(\omega_n)_{n \in \mathbb{N}}$, which forms a cover of
$\set \times \setvar$ so that every $(x,v)$ is at least in one and
in at most a finite number of ${\omega}_n$, a family of open sets
$(\mathcal{X}_n)_{n \in \mathbb{N}}$ in $\mathcal{B}(\set)$, a
sequence $(v_n)_{n \in \mathbb{N}}$ in $\setvar$ and constants
$c_n, t_n, \epsilon_n >0$, such that for $A\in\mathcal{B}(\Omega)$,
\begin{equation}\label{eq:abscont2}
 \forall (x,v) \in {\omega}_n, \forall t\in[t_n,t_n+\epsilon_n], \mathbb{P}_{(x,v)}\left((X_t, V_t) \in A \times \{v'\}\right) \geq c_n 1_{v_n=v'}\mathrm{Leb}(A\cap{\cal X}_n)
\end{equation}
which leads to bounding by below the resolvent by the following kernel
$K$ defined for $(x, v) \in \omega_n$ and $A\in\mathcal{B}(\Omega)$ as,
\begin{equation} \label{eq:bounding_kernel}
K((x,v), A \times \{v'\}) = \int \1_{A}(y) \max_{n:(x, v) \in \omega_n} \left( c_n \1_{\mathcal{X}_n\times\{v_n\}}((y,v')) \int_{t_n}^{t_n + \epsilon_n} e^{-t} \dd t \right) \dd y,
\end{equation}
and which satisfies,
\begin{equation} \label{eq:bounded_kernel}
K((x, v), \mathcal{X}_n\times\{v'\}) \geq c_n \mathrm{Leb}(\mathcal{X}_n) \int_{t_n}^{t_n + \epsilon_n} e^{-t}\dd t > 0.
\end{equation}
Thus, the kernel $K$ is a nontrivial lower semi-continuous kernel, as
shown by considering a sequence $(x_l)_l$ converging to $x$ satisfying
$K((x_l, v)), A \times \{v'\}) \geq K((x, v), A \times \{v'\})$ for
$l$ large enough. The process then is a T-process
\cite{Tuominen_1979}. As more detailed in \cite{Bierkens_2019},
another application of \eqref{eq:abscontinuity} implies that the
process is open set irreducible so that the process is a
$\phi$-process (\cite{Tweedie_1994}, Theorem 3.2). Finally, we use
also \eq{abscontinuity} to obtain the irreducibility of the
$\Delta$-skeleton chain ending the proof.

Finally, by considering a particular path, see Figure
\ref{fig:line-configuration}, we aim at getting closer to 
a uniform ergodicity property, at least for soft spheres or for a
more stringent density condition for hard spheres.

  \subsection{Proof of ergodicity} \label{ap:soft}

  The following two subsections correspond to the two main steps of
  the proof showing the ergodicity of ECMC: first we prove Lemma
  \ref{lem:submersion}, and then the irreducibility of a skeleton
  chain.

 \subsubsection{Proof of Lemma \ref{lem:submersion}}
  \label{ap:submersion}

  Let us explain briefly our strategy. First our density condition
  \eq{density2} implies roughly that we may pack 3$N$ spheres in our
  torus. We show that we can find a path between a starting and
  final configurations of $N$ spheres, furthermore proving that we
  gain \emph{density} during this path. For that, we will first
  exhibit a valid path depending on the initial and final
  configurations, starting in the situation of well-separated spheres
  and then working our way to more packed case thanks to an expansion
  procedure. We take profit of the works of \cite{Diaconis_2011} to
  construct such a procedure for a correct path and \cite{Benaim_2015}
  to get densities for such path by the notion of submersion.  In
  other words, we show how to construct a composite flow
  $\phi_{\mathbf{t}}^{\mathbf{v}}$ from
  $(x^{(0)},v^{(0)})\in \set\times\setvar$, an initial configuration,
  to $(x^{(f)},v^{(f)})\in \set\times\setvar$, a final configuration,
  where the control sequence
  $(\mathbf{t},\mathbf{v})\in\mathbb{T}_m, m\in\mathbb{N}^*,$ admits
  every pair $v=(e,i)\in\mathcal{V}$.

 \medskip
  {\noindent \bf (i)  Flows and paths.} We first define for any flow
  $\phi_{\mathbf{t}}^{\mathbf{v}}(x)$, with $\mathbf{t}=(t_i)_{i=1}^m$
  and $\mathbf{v}=(v_i)_{i=0}^m$, its corresponding cumulative time
  sequences $(T_k=\sum_{i=1}^kt_i)_{k=1}^m$ and \emph{flow path},
  i.e. a path $\gamma:[0,1]\to \Omega\times\bar{\setvar}$ so that
\begin{equation}
\gamma(s)=\begin{array}{ll}
            \left(x^{(0)} + \sum_{i=1}^{i_s} t_i \bar{v}_{i-1} + \left(sT_m-T_{i_s}\right)\bar{v}_{i_s}, \bar{v}_{i_s}\right) & \text{ for } T_{i_s} \leq sT_m < T_{i_s+1}
         \end{array}
\label{eq:path}
     \end{equation}
     with $\bar{v}_k$ defined by the mapping
     $v_k=(e=(e_0,e_1),i)\in\setvar\to \bar{v}_k= (0,\dots,e_0,
     e_1,\dots,0)\in\bar{\setvar}$, with $\bar{\setvar}$ the canonical
     basis of $\mathbb{R}^{2N}$ and $e_0$ (resp. $e_1$) placed at the
     $2i$-th (resp. $(2i+1)$-th) position. Conversely, for any path
     $\gamma:[0,1]\to \Omega\times\bar{\setvar}$ so that we can define
     sequences $\mathbf{t}=(t_i)_{i=1}^m\in\mathbb{R}_+^m$ and
     $\mathbf{v}=(v_i)_{i=0}^m\in \mathcal{V}^{m+1}$ leading to a specification
     of $\gamma$ as in \eq{path}, there is a corresponding
     flow $\phi_{\mathbf{t}}^{\mathbf{v}}$.

Now, we consider a more general path
$\gamma:[0,1]\to \Omega\times \mathbb{R}^{2N}$ so that
\begin{equation}
\gamma(s)=\begin{array}{ll}
            \left(x^{(0)} + \sum_{i=1}^{i_s} t_i n_{i-1}\bar{v}_{i-1} + \left(sT_m-T_{i_s}\right)n_{i_s}\bar{v}_{i_s}, n_{i_s}\bar{v}_{i_s}\right) & \text{ for } T_{i_s} < sT_m < T_{i_s+1}
         \end{array}
\label{eq:gen_path}
\end{equation}
with $(t_i)_{i=1}^m\in\mathbb{R}_+^m$,
$(v_i)_{i=0}^m\in \mathcal{V}^{m+1}$,
$(n_i)_{i=0}^m\in\{1\}\times\{-1,+1\}^{m-2}\times\{1\}$, then yielding
$\gamma(0),\gamma(1)\in\Omega\times\setvar$. We can define a
corresponding \emph{positive} path
$\gamma_+:[0,1]\to \Omega\times\setvar$ so that for
$(sT_m)\in
[T_{i_s},T_{i_s+1}[$ with $n_{i_s}=1$,
$\gamma_+(s) = \gamma(s)$ and for
$(sT_m)\in
[T_{i_s},T_{i_s+1}[$ with $n_{i_s}=-1$ and
$v_i = (e_i, k_i)$, we replace $\gamma$ by a continuous path
$\gamma_+$ connecting the sphere configuration reached at
$\gamma(T_{i_s}/T_m)$ to the one reached at
$\gamma(T_{i_s+1}/T_m)$ by repeating
sequential updates by at most $\epsilon>0$ increment along $+e_i$ of
all spheres but the $k_i$-th, until, by periodicity, this will amount
to an effective translation of $-t_{i+1}e_i$ of the $k_i$-th
sphere. As $\gamma(t_i)\in\Omega\times\mathbb{R}^{2N}$ is a valid
configuration with spheres separated by a pairwise distance greater
than $d_{\text{pair}}+\epsilon'$, $\epsilon'>0$, one can always consider
an increment $\epsilon=\epsilon'/2$ so that a translation of any
sphere by $+\epsilon e_i$ does not lead to any pairwise distance being
smaller than $d_{\text{pair}}+\epsilon'/2$, making the positive path a valid one,
i.e. $\gamma_+:[0,1]\to\Omega\times\bar{\setvar}$. We then can define
the flow $\phi_{\mathbf{t}_+}^{\mathbf{v}_+}$ corresponding to the
positive path $\gamma_+$.

Thus, finding a composite flow $\phi_{\mathbf{t}}^{\mathbf{v}}$ from
$(x^{(0)},v^{(0)})$ to $(x^{(f)},v^{(f)})$ amounts to finding a path
$\gamma:[0,1]\to\Omega\times\mathbb{R}^{2N}$ so that
$\gamma(0) = (x^{(0)},v^{(0)})$ and $\gamma(1)=(x^{(f)},v^{(f)})$. We
introduce $u_{\mathsf{x}} = (1,0)$ and $u_{\mathsf{y}} = (0,1)$, the unitary vectors aligned
with the $x$-axis and $y$-axis respectively. Without loss of
generality and out of simplicity, we set $v^{(0)} = (u_{\mathsf{x}},1)$ and
$v^{(f)} \neq (u_{\mathsf{y}}, N)$, as a different setting only impacts the
construction order of $\mathbf{t}$ and $\mathbf{v}$.

\medskip
{\bf  \noindent (ii) Connectivity in the fully-expanded case.} For $(x,x')\in \Omega^2$, we define,

\begin{equation}
  I(x, x') =\min_{i\neq j}d(x_i,x'_j),
  \label{eq:I}
\end{equation}
the minimal distance between any two spheres respectively picked in $x$ and $x'$.

We first consider the case where $(x^{(0)},x^{(f)})\in\Omega ^2$ is
such that
\begin{equation}
  I(x^{(0)},x^{(0)})> 2d_{\text{pair}}, I(x^{(f)},x^{(f)})> 2d_{\text{pair}} \text{ and } I(x^{(0)},x^{(f)})> 2d_{\text{pair}}.
  \label{eq:sep}
  \end{equation}
  Considering the first sphere initially positioned in $x^{(0)}_1$, we
  consider the continuous flow
  $\phi_{\mathbf{\tilde{t}}_1}^{\mathbf{\tilde{v}}_1}$ set by
$$\mathbf{\tilde{t}}_1 = 
\{t_{u_{\mathsf{x}}}, t_{u_{\mathsf{y}}}(x^{(f)}_{1,0}-x^{(0)}_{1,0}) \mod L,(x^{(f)}_{1,1}-x^{(0)}_{1,1})
\mod L\} \text{ and } \mathbf{\tilde{v}}_1 = \{u_{\mathsf{x}},u_{\mathsf{y}}\}$$
with
$$ t_{u_{\mathsf{x}}}=\left\{\begin{array}{l}
    (x^{(f)}_{1,0}-x^{(0)}_{1,0}) \mod L  \text{ if } |x^{(f)}_{1,0}-x^{(0)}_{1,0}| >0       \\
      L  \text{ otherwise}
    \end{array}\right.
  \qquad
  t_{u_{\mathsf{y}}}=\left\{\begin{array}{l}
    (x^{(f)}_{1,1}-x^{(0)}_{1,1}) \mod L  \text{ if } |x^{(f)}_{1,1}-x^{(0)}_{1,1}| >0       \\
      L  \text{ otherwise}
    \end{array}\right.\
$$
The flow $\phi_{\mathbf{\tilde{t}}_1}^{\mathbf{\tilde{v}}_1}$ pushes
$x^{(0)}$ to $(x^{(f)}_1,x^{(0)}_2,\dots,x^{(0)}_N)$ and admits
$(u_{\mathsf{x},1)}$ and $(u_{\mathsf{y}},1)$ in $\mathbf{\tilde{v}_1}$ over strictly
 positive times. We now need to modify the sequences
$\mathbf{\tilde{t}}_1$ and $\mathbf{\tilde{v}}_1$ into sequences
$\mathbf{t}_1$ and $\mathbf{v}_1$ so that the minimal pairwise
distance along $\phi_{\mathbf{t}_1}^{\mathbf{v}_1}$ is strictly
greater than $d_{\text{pair}}$.

To do so, we consider the corresponding path
$\tilde{\gamma}_1:[0,1]\to\set\times\bar{\setvar}$ to the flow
$\phi_{\mathbf{\tilde{t}}_1}^{\mathbf{\tilde{v}}_1}$ and we modify it
into a path $\gamma_1:[0,1]\to\set\times\mathbb{R}^{2N}$. As the path
$\tilde{\gamma}_1$ is continuous and using condition \eq{sep}, the
collection of times $s\in[0,1]$ at which the distance constraint is
not satisfied along $\tilde{\gamma}_1$ can be written down as a
disjoint union of intervals $S=\cup_{k=1}^K[a_k,b_k]$ and so that
there is only one sphere $i_k$ verifying
$d(\tilde{x}_{1} - x^{(0)}_{i_k})\leq d_{\text{pair}}$ for
$s \in [a_k,b_k]$ and $\tilde{\gamma}(s)=(\tilde{x},\tilde{v})$,
$\tilde{v}$ updating the first sphere. Now, for $s\not \in S$, we set
$\gamma_1(s) = \tilde{\gamma}_1(s)$ and for $s \in [a_k,b_k]$, we
modify $\tilde{\gamma}_1$ into a continuous path $\gamma_1$ connecting
$\tilde{\gamma}_1(a_k)$ and $\tilde{\gamma}_1(b_k)$ so that it is
composed of moves of the first sphere along $\pm u_{\mathsf{x}}$ and $\pm u_{\mathsf{y}}$
and verifies
$d_{\text{pair}} +\epsilon_1/2 < d(x_{1}, x^{(0)}_{i_k}) <
d_{\text{pair}} + \epsilon_1, \epsilon_1 > 0$ for $s\in[a_k,b_k]$ and
$\gamma_1(s)=(x,v)$. As
$I(x^{(0)}, x^{(0)}) > 2 d_{\text{pair}}$, we can choose
$\epsilon_1 > 0 $ so that
$I(x,x^{(0)}) > d_{\text{pair}}+\epsilon_1/2$ for any $x\in\Omega$ reached by
$\gamma_1(s)$ with $s\in[a_k,b_k]$ and $k\in\{1,\dots,K\}$.

Eventually, the modified path $\gamma_1$ is continuous, connects
$(x^{(0)}, v^{(0)})$ and
$((x^{(f)}_1,x^{(0)}_2,\dots,x^{(0)}_N), u_{\mathsf{y}})$ and respects a minimal
pairwise distance of $d_{\text{pair}} + \epsilon_1/2$. Up to the
definition of a corresponding positive path, it yields a flow
$\phi_{\mathbf{t}_1}^{\mathbf{v}_1}$ updating the sphere $1$ to its
final position while respecting a minimal pairwise distance greater
than $d_{\text{pair}}+\epsilon_1/4$. The complete flow
$\phi_{\mathbf{t}}^{\mathbf{v}}$ is then obtained by iteration of this
procedure for each sphere and the composition of the respective flow
$\phi_{\mathbf{t}_i}^{\mathbf{v}_i}$, which keeps a minimal pairwise
distance of $d_{\text{pair}}+\min_{i\in\setind} \epsilon_i/4$. Thus,
the flow $\phi_{\mathbf{t}}^{\mathbf{v}}$ connects $(x^{(0)},v^{(0)})$
to $(x^{(f)},v^{(f)})$, admits every $(e,i)\in\setvar$ in $\mathbf{v}$
and keeps a minimal pairwise distance strictly greater than
$d_{\text{pair}}$.

\medskip
{\bf \noindent (iii) Connectivity in the individually-expanded case.} We now consider the case where
$(x^{(0)},x^{(f)})\in\Omega ^2$ is such that
$$I(x^{(0)},x^{(0)})> 2d_{\text{pair}}, I(x^{(f)},x^{(f)})> 2d_{\text{pair}} \text{ and } I(x^{(0)},x^{(f)}) \leq  2d_{\text{pair}}.$$
Finding a path connecting such configurations is immediate through the procedure {\bf (ii)} if we can
consider an intermediate configuration $x^{(I)}\in \Omega$ so that,
$$I(x^{(I)},x^{(I)})> 2d_{\text{pair}}, I(x^{(0)},x^{(I)})> 2d_{\text{pair}} \text{ and } I(x^{(I)},x^{(f)}) >  2d_{\text{pair}}.$$
As noted in \cite{Diaconis_2011} for a more stringent density
constraint, such a configuration $x^{(I)}$ is easily constructed by
induction given the density constraint in lemma \ref{lem:submersion},
since, for any $(x^{(0)},x^{(f)},x^{(I)})\in \Omega^3$, we have
$\text{Vol}(\cup_{i=1}^NB(x^{(0)}_i,
d_{\text{pair}})\cup_{i=1}^NB(x^{(f)}_i,
d_{\text{pair}})\cup_{i=1}^NB(x^{(0)}_i, d_{\text{pair}})) \leq 3N \pi
d_{\text{pair}}^2 < \frac{\pi L^2\sqrt{3}}{6}$, where $\frac{\pi\sqrt{3}}{6}$ is the highest
achievable density of a disk packing into a periodic square box.

\medskip
{\bf\noindent (iv) Connectivity in the collapsed case (expansion/collapse procedure).}  We now consider the case where
$(x^{(0)},x^{(f)})\in\Omega ^2$ is such that
$$d_{\text{pair}}<I(x^{(0)},x^{(0)})\leq 2d_{\text{pair}} \text{ or } d_{\text{pair}}<I(x^{(f)},x^{(f)})\leq 2d_{\text{pair}}.$$
Building on the procedures {\bf (ii)} and {\bf (iii)}, we only need to
show how to find a path connecting $x^{(0)}$ to a configuration
$x^{(I)}$ so that $I(x^{(I)},x^{(I)})> 2d_{\text{pair}}$. The case
where $I(x^{(f)},x^{(f)})\leq 2 d_{\text{pair}}$ can be dealt with by
showing the possibility of a path connecting $x^{(f)}$ to a
configuration $x'^{(I)}, I(x'^{(I)},x'^{(I)})>2d_{\text{pair}}$,
reversing the path and taking its positive counterpart.

Adapting the expansion procedure in \cite{Diaconis_2011}, it comes down to proving
that there exists some $\delta>0$ so that, for any $(x,v)\in \Omega\times\setvar$ with
$I(x,x) < 2d_{\text{pair}}$, there exists a path $\gamma_\delta$ such
that $\gamma_\delta(0)=(x,\bar{v})$, $\gamma_\delta(1)=(x',\bar{v}')$ and
$I(x',x')\geq I(x,x) + \delta$. Indeed,
assuming such a $\delta>0$ exists, we consider the maximal pairwise
distance $M$ achievable from $x^{(0)}$ by a path $\gamma$, i.e.
$$M = \max_{y\in \mathcal{I}(x^{(0)})} I(y,y) \text{ with } \mathcal{I}(x^{(0)})=\{y\in\Omega; \exists \gamma\in C([0,1],\Omega\times\bar{\setvar}), \gamma(0)=(x^{(0)},\bar{v}^{(0)}), \gamma(1) = (y,\bar{v}_y)\},$$
As $I$ is a bounded function, $M$ is finite and given
$\eta\in]0,\delta/2[$, there exists $y_1\in\mathcal{I}(x^{(0)})$ such that
$I(y_1,y_1) \geq M - \eta$. If $I(y_1,y_1)<2d_{\text{pair}}$, we can
consider a path $\gamma_\delta$ so that $\gamma_\delta(0)=(y_1,\bar{v})$ and
$\gamma_\delta(1)=(y_2,\bar{v}')$ and $I(y_2,y_2) \geq I(y_1,y_1) + \delta$. By
construction, $y_2\in\mathcal{I}(x^{(0)})$ and $I(y_2,y_2) > M$, which is
impossible. It then shows that there exists $y\in\mathcal{I}(x^{(0)})$ such
that $I(y,y)>2d_{\text{pair}}$. It leads to the construction of an
intermediate configuration $x^{(I)}$, with
$I(x^{(I)},x^{(I)})>2d_{\text{pair}}$ with a valid flow starting from
$x^{(0)}$, and conversely by reverting the flow one may then find a
flow to collapse the configuration into $x^{(f)}$.

Starting from $(x^{(0)},v^{(0)})$, the system evolves to $(x,v^{(0)})$
until a first refreshment time $t_0$ updating the state to
$(x,v)\in \Omega\times\setvar$
Let us now prove that there exists some $\delta>0$ so that, for any
$x\in \Omega$ with $I(x,x) < 2d_{\text{pair}}$, there exists a path
$\gamma_\delta$ and $v\in\setvar$ such that
$\gamma_\delta(0)=(x,\bar{v})$, $\gamma_\delta(1)=(x',\bar{v}')$ and
$I(x',x')\geq I(x,x) + \delta$. To do so, we will use the induction
strategy of the proof of \cite[Lem. 4.2]{Diaconis_2011}, and adapt it
here to the PDMP scheme and torus setting.

We proceed by induction and thus have to prove
that there exists $\delta=\delta(N, d_{\text{pair}}, L)$ and
$\rho=\rho(N, d_{\text{pair}}, L)$ such that
$\forall J\in\{1,\cdots,N\}$,
$$ (P(J))\quad\begin{array}{l}
\forall x\in\Omega(J)\,s.t.\,
I(x,x)<2d_{\text{pair}},\\
\qquad \exists v\in\setvar,\exists\gamma\in C([0,1],\Omega(J)\times\bar{\setvar}),\mbox{
  with }\left\{\begin{array}{l}\gamma(0)=(x,\bar{v}),\,\gamma(s)=(x(s),\bar{v}(s)),\\ I(x(1),x(1))\ge I(x,x)+\delta,\\
   \sup_t\max_{i}
d(x_i,x(t)_i)\le J\rho
\end{array}\right.\end{array}$$ where we remind that
$$  \Omega(J)=\{x\in(\mathbb{R}/L\mathbb{Z})^{2J};\forall
(i,j)\in \llbracket 1, J \rrbracket^2,i\neq j, (x_i,x_j)\in\Omega_{\text{pair}}\},
$$
stands for the system with only $J$ spheres. 
Let us choose $\delta =d_{\text{pair}}/(600N^2)$ and
$\rho~=~d_{\text{pair}}/(6N)$.

Let us start our induction with $J=1$ (only one sphere) for which
there is nothing to prove. Suppose now that $P(J-1)$ is verified. Let
us divide, as equivalence class, spheres that can be connected by a
path in $\cup_1^J B(x_k, d_{\text{pair}}+J\rho)$. If there are more
than two classes, then one can consider each class individually, which
then contains strictly less than $N$ spheres, and use the induction
hypothesis. Indeed, in this case one has
$\{1,\cdots N\}=I\cup \tilde I$ and for all $i\in I,\, j\in \tilde I$
one has $d(x_i,x_j)>2d_{\text{pair}}+2J\rho$. The induction hypothesis
enables us to build two paths $\gamma_I$ and $\gamma_{\tilde I}$ each
satisfying $(P(I))$ and $(P(\tilde I))$ and this defines a path
$\gamma$ for all spheres. As along $\gamma_I$,
$\sup_t \max_i d(x_i,x(t)_i)\leq |I|\rho$ (and respectively along
$\gamma_{\tilde{I}}$), we have to impose here that
$2d_{\text{pair}}+2J\rho\delta-(|I|+|\tilde
I|)\rho>2d_{\text{pair}}+\delta$ resuming in $J\rho>\delta$, to get
that $\gamma$ is a valid path and $I(x(1),x(1))>I(x,x)+\delta$,
$\sup_t \max_id(x_i,x(t)_i)<\max(|I|,|\tilde I|)\rho\le (J-1)\rho$.

Thus we are reduced to consider the case where there is only one
equivalence class, so that the spheres are quite packed. Recall that
our density assumption ensures that there is in fact sufficient space
for at least $3N$ spheres of radius $d_{\text{pair}}$. One can then
choose a direction $\nu=(1,0)$ or $(0,1)$ and introduce an ordering
$\sigma:\setind\to\setind$ so that the reordered positions
$\{x_{\sigma(i)}\}_{i=1}^J$ satisfy
$\langle x_{\sigma(i)},\nu\rangle\leq \langle
x_{\sigma(j)},\nu\rangle$ for all $(i,j)\in\setind^2$ with
$\sigma(i)<\sigma(j)$. As
$\{\langle x_{\sigma(i)},\nu\rangle\}_{i=1}^J\in [0,L]^J$, we can now
consider a pair of furtherst apart $\sigma-$successive spheres along
the $\nu$-direction, i.e. a pair of spheres $i_J$ and $i_1$ so that
$\sigma(i_J) = \left(\sigma(i_1) + 1\right) \mod J$ and
    $$\left(\langle x_{i_J}-x_{i_1}, \nu \rangle\right)\mod L =
    \max_{i}\left(\langle \left(x_{\sigma^{-1}\left(\sigma(i)+1 \text{
              mod} J\right)}-x_{i}\right), \nu \rangle\mod L\right),$$
    which is, for a well-chosen direction $\nu$, larger than
    $\max \{d\in ]d_{\text{pair}},+\inf[;J \leq \lfloor L/d\rfloor
    \lfloor (L/d) (2\sqrt{3}/3)\rfloor \}$ which is in turn larger
    than
    $\max \{d\in ]d_{\text{pair}},+\inf[; N \leq \lfloor L/d\rfloor
    \lfloor (L/d)(2\sqrt{3}/3)\rfloor \}\geq 2
    (2d_{\text{pair}}/\sqrt{3})$. Indeed, given the density condition,
    one can pack $3N$ spheres of radius $d_{\text{pair}}$. Considering
    the densest hexagonal packing of these $3N$ spheres, we can pack
    $N$ spheres of radius $(2d_{\text{pair}}/\sqrt{3})$ corresponding
    to the circumscribed circles of the equilateral triangles forming
    up the $3N$-hexagonal packing.

    We then note the subset of indices matching $i_J$ in its
    $\nu$-coordinate
    $\mathcal{I}_J=\{i\in\setind\setminus\{i_J\}; \langle
    x_{i},\nu\rangle = \langle x_{i_J},\nu\rangle\}.$ From there, we
    build on the ordering $\sigma$ to obtain a sequence
    $(i_l)_{l=1}^{J}$ with $i_J$ and $i_1$ as previously defined and
    $i_l$ so that
    $\sigma(i_l) = \left(\sigma(i_{l-1}) - 1\right) \mod J$ for
    $2\leq l \leq J-1-|\mathcal{I}_1|$ and $i_l\in \mathcal{I}_J$ for
    $J>l\geq J-|\mathcal{I}_J|$.
    Now, for $j\in\{1,...,J\}$, choose $a_j= (J + 1 - j)\rho$
    and consider $v=(\nu, i_1)$ and the valid continuous path $\gamma$ set by the
    sequences
  \begin{equation} 
    \begin{aligned}
    &\mathbf{t} = \left(a_{l}\right)_{l=1}^{J}, \;
    \mathbf{v} = \left((\nu,i_l) \right)_{l=1}^{J}
    \label{eq:sets_for_exp}
     \end{aligned}
\end{equation}
It is easy to verify as in \cite{Diaconis_2011} that in this case
$(P(J))$ is verified. Indeed, we first have that\\
$\sup_t\max_i d(x_i,x(t)_i) \leq J\rho$. Note that there is no
periodicity effect to take care of here, as $J\rho < L/2$ since
$L^2 > 3N\pi d_{\text{pair}}^2>(d_{\text{pair}}/3)^2$.

Then, we consider the evolution of $d(x_{i_l}(t),x_{i_m}(t))$ for all
$1\leq l < m \leq J$. The periodic distance $d$ refers to the shortest
distance between every periodic copies of the spheres $i_l$ and
$i_m$. We refer to the initially involved copies as
$x^c_{i_l}$ and $x^c_{i_m}$. A change of the copy involved in the
computation of $d$ means that the relative displacement
$(m-l)\rho > L/2 - |\langle x^c_{i_l}-x^c_{i_m} ,
\nu\rangle|$. However, as $N\rho < L/2 - 2 d_{\text{pair}} -\delta$,
it means that, first the initial periodic distance is bigger than
$2d_{\text{pair}}>I(x,x)$ and second that the final periodic distance
is
$L-(|\langle x^c_{i_l}-x^c_{i_m} , \nu\rangle| +(m-l)\rho) >
2d_{\text{pair}} + \delta > I(x,x) +\delta$. 

Now, for pairs of spheres $2\leq l < m < J-|\mathcal{I}_1|$ with
unchanged involved periodic copies, we have,
\begin{eqnarray*}
|(x^c_{i_l}+a_{l}\nu) - (x^c_{i_m}+a_{m}\nu)|^2&=& |x^c_{i_l}-x^c_{j_l}|^2+2(a_{l}-a_{m})\langle x^c_{i_l}-x^c_{i_m},\nu\rangle +|a_{l}-a_{m}|^2 \ge |x^c_{i_l}-x^c_{i_m}|^2+\rho^2.
\end{eqnarray*}
For pairs of spheres $J - |\mathcal{I}_J| \leq l<m\leq J$, we have,
\begin{eqnarray*}
|(x^c_{i_l}+a_{l}\nu) - (x^c_{i_m}+a_{m}\nu)|^2&=& |x^c_{i_l}-x^c_{i_m}|^2+|a_{l}-a_{m}|^2 \ge |x^c_{i_l}-x^c_{i_m}|^2+\rho^2.
\end{eqnarray*}
And finally for $J - |\mathcal{I}_J| \leq l<m\leq J$ and
$2\leq  m < J-|\mathcal{I}_1|$, with
unchanged involved periodic copies, we have,
\begin{eqnarray*}
|(x^c_{i_l}+a_{l}\nu) - (x^c_{i_m}+a_{m}\nu)|^2&=& |x^c_{i_l}-x^c_{i_m}|^2+2(a_{l}-a_{m})\langle x^c_{i_l}-x^c_{i_m},\nu\rangle +|a_{l}-a_{m}|^2 
\end{eqnarray*}
If $\langle x^c_{i_l}-x^c_{i_m},\nu\rangle < 0$, then
$$
|(x^c_{i_l}+a_{l}\nu) - (x^c_{i_m}+a_{m}\nu)|^2\ge |x^c_{i_l}-x^c_{i_m}|^2 +\rho^2.
$$
Otherwise,  
\begin{equation*}
\begin{split}
  |(x^c_{i_l}+a_{l}\nu) - (x^c_{i_m}+a_{m}\nu)|^2\ge|\langle (x^c_{i_J}+a_{J}\nu) - (x^c_{i_1}+a_{1}\nu), \nu \rangle|^2
 \\\geq
 \left(\frac{4}{\sqrt3}d_{\text{pair}} - (J-1)\rho\right)^2 
 > 4d^2_{\text{pair}} + \rho^2>I(x,x)^2 + \rho^2
\end{split}
\end{equation*}

We then deduce $I(x(1),x(1))^2\ge I(x,x)^2+\rho^2\ge (I(x,x)+\delta)^2$ .

Remark that using the reversed-path scheme, as described in the
paragraph {\bf (i)}, we also have a valid collapse scheme. Therefore using the
expansion then collapse procedure we may go from a collapsed initial
configuration to an intermediate separated configuration by expansion
then move to an other intermediate separated configuration and then
use a collapse procedure (if needed) to reach the final configuration.

\medskip {\bf \noindent (v) Gaining density.} Considering any
$(x^{(0)},v^{(0)}),(x^{(f)},v^{(f)})\in\set\times\setvar$, there
exist admissible deterministic sequences
$(\mathbf{t}=(t_i)_{i=1}^m,\mathbf{v}=(v_i)_{i=0}^m)\in\mathbb{T}_m$,
$m\in\mathbb{N}^*$, so that the corresponding flow
$\phi_{\mathbf{t}}^{\mathbf{v}}$ pushes $(x^{(0)},v^{(0)})$ to
$(x^{(f)},v^{(f)})$ and $\mathbf{v}$ admits every pair in $\setvar$ at
least once. Recalling the notation $T_k = \sum_{i=1}^{k}t_k$, we
consider a bounded neighbourhood $\mathcal{U}_k$ of $T_k$ for
$1\leq k \leq m$. The neighbourhoods $(\mathcal{U}_k)_{k=1}^{m}$ can
be chosen so that they do not intersect and so that, for any sequences
$(\bm{\tau},\mathbf{v})\in\mathcal{T}$, with
$\mathcal{T}=\{\bm{s}\in\mathbb{T}_m;
\sum_{l=1}^{k}s_l\in\mathcal{U}_k\}$, the flow
$\phi_{\bm{\tau}}^{\mathbf{v}}$ preserves a minimal pairwise
interdistance strictly greater than $d_{\text{pair}}$, as the initial
flow $\phi_{\mathbf{t}}^{\mathbf{v}}$ already does.

Following \cite[section 6]{Benaim_2015} and as done in the proof of
lemma 8 in \cite{Bierkens_2019}, we only need to show that, for some
$t'>T_m$, the partial map
$\bm{\tau}\to \phi_{t' - \sum_{k=1}^m\tau_m}^{v^{(f)}} \circ
\phi_{\bm{\tau}}^{\mathbf{v}}(x^{(0)})$, defined on
$\bm{\tau}\in\mathcal{T}, \sum_{k=1}^m\tau_k<t'$, has full rank with
$$ \phi_{t' - \sum_{k=1}^{m}\tau_k}^{v^{(f)}} \circ \phi_{\bm{\tau}}^{\mathbf{v}}(x^{(0)}) = x^{(0)} + \tau_1 \bar{v}_1 + \tau_2 \bar{v}_{2}+\dots+\tau_m \bar{v}_{m} + (t'-\tau_{m}) \bar{v}^{(f)}.$$
The image of the differential of the partial mapping is spanned by the
vector family,
$$\{g_i\}_{i=1}^{m}=  \left\{(\bar{v}_{1}-\bar{v}_2), (\bar{v}_{2}-\bar{v}_3),\dots,(\bar{v}_{m}-\bar{v}^{(f)})\right\}.
$$
As shown by the construction in the procedure {\bf (ii)}, the family
$\{g_i\}_{i=1}^{m}$ includes the following family of $2N$ vectors,
$$\left\{(\bar{v}_{\mathsf{x},1}-\bar{v}_{\mathsf{y},1}), (\bar{v}_{\mathsf{y},1}-\bar{v}_{\mathsf{x},2}), (\bar{v}_{\mathsf{x},2}-\bar{v}_{\mathsf{y},2}),\dots,(\bar{v}_{\mathsf{x},N}-\bar{v}_{\mathsf{y},N}),(\bar{v}_{\mathsf{y},N}-\bar{v}^{(f)})\right\},
$$
with $\bar{v}_{\mathsf{x},i}, \bar{v}_{\mathsf{y},i}$ corresponding to
$(u_{\mathsf{x}},i),(u_{\mathsf{y}},i)$ up to the $v\to \bar{v}$
mapping defined in {\bf (i)}. Thus, the family $\{g_i\}_{i=1}^{m}$
spans $\Omega$, as one can reconstruct the basis
$\{\bar{v}_{\mathsf{x},i}\}^{N}_{i=1}\cup\{\bar{v}_{\mathsf{y},i}\}^{N}_{i=1}$
from this family by composition since
$\bar{v}_{\mathsf{y},N}\neq\bar{v}^{(f)}$ and $\bar{v}^{(f)}$ appears
elsewhere in $\{g_i\}_{i=1}^{m}$.

The partial mapping
$\bm{\tau}\to\phi_{t' - \sum_{k=1}^{m}\tau_k}^{v^{(f)}} \circ
\phi_{\bm{\tau}}^{\mathbf{v}}(x^{(0)})$ then has full rank and is a
submersion at $\mathbf{t}$, proving Lemma
\ref{lem:submersion}.

\subsubsection{Irreducibility of the $\Delta$-skeleton}
\label{ap:irr}

    \begingroup
    \allowdisplaybreaks

    The $\Delta$-chain $(X_{k\Delta}, V_{k\Delta})_{k \in \mathbb{N}}$
    of the process, or its skeleton chain, is the sampled chain at
    times $t_k = k\Delta$ with $k \in \mathbb{N}$ and
    $\Delta \in \mathbb{R}_+^*$.  It is irreducible if for all
    $(x, v) \in \set \times \setvar$ and
    $(y, v') \in \set \times \setvar$, there exist a neighbourhood
    $\mathcal{Y}$ of $y$ and $n \in \mathbb{N}$ such that
    $\mathbb{P}_{(x, v)}((X_{n\Delta}, V_{n\Delta}) \in \mathcal{Y} \times
    \{v'\}) > 0$.

    Let $((x, v), (y, v')) \in (\set \times \setvar)^2$ and let
    $(w, u) \in \set \times \setvar$. Applying Lemma
    \ref{lem:submersion} between the pairs $((x, v), (w, u))$,
    $((w, u), (y, v'))$ and $((w, u), (w, u))$ (looping around) yields
    that there exists $\epsilon \in \mathbb{R}_+^*$ and $\tau, c > 0$
    such that for any $t \in[\tau, \tau+\epsilon]$, there exists a
    neighbourhood $\mathcal{Y}$ of $y$ such that,
    \begin{equation}\label{eq:lemma_appli}
      \mathbb{P}_{(x, v)}((X_{t}, V_t) \in \cdot \times \{v'\}) \geq c\text{Leb}(\cdot \cap \mathcal{Y}) > 0.
    \end{equation}
    While $\tau$ and $c$ depends on $(x,v), (y,v')$ and $(w,u)$,
    $\epsilon$ can be set so as to depend only $(w,u)$ and the ability
    for the process to loop around an arbitrary point. We refer to
    \cite{Bierkens_2019} for the successive applications of Lemma
    \ref{lem:submersion} and corollary \eq{abscontinuity} resulting in
    \eqref{eq:lemma_appli}.
    
    Finally, setting $\Delta = \epsilon$ and
    $n = \left\lfloor \frac{\tau}{\epsilon} + 1 \right\rfloor$
    such that
    $n \epsilon \in [\tau,\tau+\epsilon]$, we obtain
    $\mathbb{P}_{(x,v)}(X_{n\Delta}, V_{n\Delta}~\in~\cdot~\times~\{v'\}) > 0$.
 \endgroup
   
    \subsection{Towards uniform ergodicity in the soft-sphere case}
    \label{ap:quant}

    \begingroup
    \allowdisplaybreaks

\begin{figure}[t]
  \centering
  \includegraphics[width=1.0\textwidth]{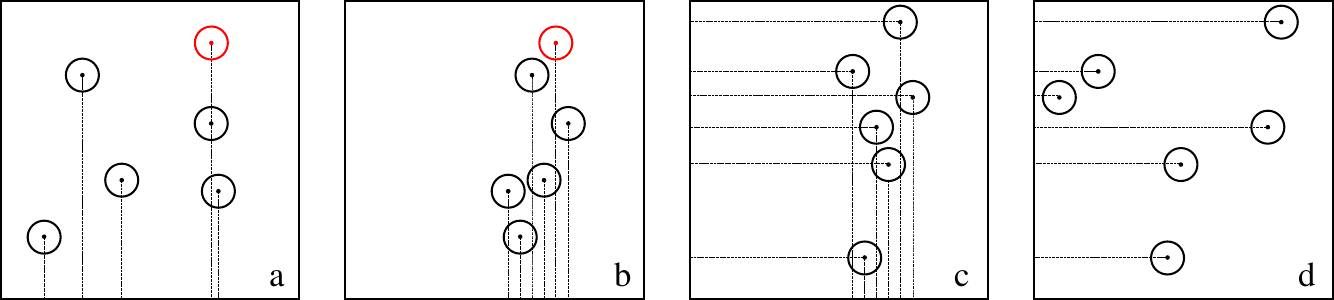}
  \caption{In the case $\alpha>0$, to reach the configuration \textit{(d)} from any initial
    configuration \textit{(a)}, the x-coordinates of the spheres are
    first gathered in configuration {\it (b)} in a segment of size
    $\frac{L}{N}$ with $L$ the size of the box side and $N$ the number
    of spheres \textit{(b-d)}. They are then moved successively along
    their $y$-coordinates until they reach their final $y$-position in
    configuration {\it (c)}. From {\it (a)} to {\it (b)} and from {\it
      (b)} to {\it (c)}, a distance of $\frac{L}{N^2}$ in the
    $x$-coordinates is preserved between any pair of spheres. They are
    finally updated to their final $x$-position in the configuration
    {\it (f)}, while a distance of $\alpha$ is preserved in the
    $y$-coordinates. }
  \label{fig:line-configuration}
\end{figure}

Let us see now how we can consider a modification of the reachability
procedure described in {\bf (ii)}, so that the initial condition
dependence only appears in travel times. It may be seen as a first
step towards uniform ergodicity and coupling. The described procedure does not
behave optimally in the number of spheres and, consequently we only
describe it qualitatively. The notations of the previous subsection
apply and we consider the soft-sphere case ($d_{\text{pair}} =
0$). The same could be done in the hard-sphere case however with a far
more stringent density condition than \eq{density2}.

    We introduce the quantity,
    $$\alpha = \max_{k\in\{0,1\}}\min_{i\neq j}\left(\min(|x_{i,k}^{(f)}-x^{(f)}_{j,k}|, L - |x_{i,k}^{(f)}-x^{(f)}_{j,k}|)\right)$$
    and consider out of simplicity that $\alpha$ is reached along the
    y-coordinate ($k=1$) as, otherwise, the following points still
    hold, up to exchanging $k=0$ and $k=1$. Note that $v^{(f)}$ is not
    really important as we consider a final refreshment to reach
    it. The followed strategy is illustrated in Fig.~\ref{fig:line-configuration}.
    
    Starting from $(x^{(0)},v^{(0)})$, the system evolves to
    $(\tilde{x}^{(0)},\tilde{v}^{(0)})$ until a first refreshment
    time. There, we consider the same ordering $\sigma$ along the
    $x$-coordinate as described in {\bf (iv)} to determine the
    furthest apart successive spheres along the $x$-axis. We note
    $i_1$ and $i_N$ the indices of these spheres, so that
    $\left(\tilde{x}^{(0)}_{i_N,0}-\tilde{x}^{(0)}_{i_1,0}\right)\mod
    L$ is either $0$, in the case where all x-coordinates are the same
    in $\tilde{x}^{(0)}$, or larger than $L/N$ otherwise. We then note
    the subset of indices matching $i_N$ in its x-coordinate
    $\mathcal{I}_N=\{i\in\setind\setminus\{i_N\};
    \tilde{x}^{(0)}_{i,0} = \tilde{x}^{(0)}_{i_N,0}\}.$ We obtain a
    sequence $(i_k)_{k=1}^N$ where $i_1$ and $i_N$ as previously
    defined, $i_l$ so that $\sigma(i_l)=(\sigma(i_{l-1}-1))\mod N$ for
    $2\leq l \leq N - 1 -|\mathcal{I}_N|$ and $i_l\in \mathcal{I}_N$ otherwise.

    We now \textit{stack} the $x$-coordinates of the spheres on a
    segment
    $$\left[x_{i_N,0}^{(0)} -(N-|\mathcal{I}_N|)L/N^2, x_{i_N,0}^{(0)}+
    |\mathcal{I}_N|L/N^2\right]/(L\mathbb{Z}),$$ by
    following the trajectories defined by the sequences,
  \begin{equation}
    \begin{aligned}
    &\mathbf{t}_{\mathsf{x}} = \left(\left(\tilde{x}^{(0)}_{i_{N},0} - \tilde{x}^{(0)}_{i_{l},0}\right)\mod L-\frac{lL}{N^2}\right)_{l=1}^{N-1-|\mathcal{I}_N|}, \;
    \mathbf{v}_{\mathsf{x}} = \left((u_{\mathsf{x}},i_l) \right)_{l=1}^{N-|\mathcal{I}_N|-1}\\
    &\mathbf{t}^{\mathcal{I}_N}_{\mathsf{x}} = \left(\frac{(|\mathcal{I}_N|-l)L}{N^2}\right)_{l=0}^{|\mathcal{I}_N|-1}, \;
    \mathbf{v}_{\mathsf{x}}^{\mathcal{I}_N} = \left((u_{\mathsf{x}},i_{l}) \right)^{N-1}_{l=N-|\mathcal{I}_N|}.
     \label{eq:sets_for_line}
     \end{aligned}
\end{equation}
The corresponding flow
$\phi_{\mathbf{t}_{\mathsf{x}}^{\mathcal{I}_N}}^{\mathbf{v}_{\mathsf{x}}^{\mathcal{I}_N}}\circ\phi_{\mathbf{t}_{\mathsf{x}}}^{\mathbf{v}_{\mathsf{x}}}(\tilde{x}^{(0)})=x^{(\mathsf{x})}$
exists as it preserves a bound for all pair event rates
$\lambda_{j\neq i_l}(x,(u_{\mathsf{x}},i_l)) \leq\lambda_{i_{l-1}}(x,(u_{\mathsf{x}},i_l))  \leq
\gamma\sigma^\gamma/(L/N^2)^{\gamma+1}$, thanks to the $(i_l)_{l=1}^N$ ordering.

{\noindent \bf Case $\alpha>0$.} From the configuration
$x^{(\mathsf{x})}$, we then updates the y-coordinates of the spheres
to their final values by introducing the following sequences,
  \begin{equation}
    \mathbf{t}_{\mathsf{y}} = \left(\left(x^{(f)}_{i,1} - \tilde{x}^{(0)}_{i,1}\right)\mod L\right)_{i=1}^{N}, \;
     \mathbf{v}_{\mathsf{y}} = \left((u_{\mathsf{y}},i) \right)_{i=1}^{N},
    \label{eq:sets_for_line}
\end{equation}
and the corresponding flow
$\phi_{\mathbf{t}_{\mathsf{y}}}^{\mathbf{v}_{\mathsf{y}}}(x^{(\mathsf{x})})=x^{(\mathsf{y})}$
exists as it preserves the same bound by keeping a minimum pairwise
distance of at least $L/N^2$. 

Eventually, the sphere x-coordinates
are updated to their final values by considering the sequences,
  \begin{equation}
    \mathbf{t}_{\mathsf{f}} = \left(\left(x^{(f)}_{i,0} - x^{(\mathsf{x})}_{i,0}\right)\mod L\right)_{i=1}^{N}, \;
     \mathbf{v}_{\mathsf{f}} = \left((u_{\mathsf{x}},i \right)_{i=1}^{N}, v^{(f)}),
    \label{eq:sets_for_line}
\end{equation}
and the corresponding flow
$\phi_{\mathbf{t}_{\mathsf{f}}}^{\mathbf{v}_{\mathsf{f}}}(x)=x^{(f)}$
exists as it preserves a minimum pairwise distance of at least
$\alpha$ and the total composite flow
$\phi_{\mathbf{t}}^{\mathbf{v}}=\phi_{\mathbf{t}_{\mathsf{f}}}^{\mathbf{v}_{\mathsf{f}}}\circ\phi_{\mathbf{t}_{\mathsf{y}}}^{\mathbf{v}_{\mathsf{y}}}\circ\phi_{\mathbf{t}_{\mathsf{x}}^{\mathcal{I}_f}}^{\mathbf{v}^{\mathcal{I}_N}_{\mathsf{x}}}\circ\phi_{\mathbf{t}_{\mathsf{x}}}^{\mathbf{v}_{\mathsf{x}}}$,
$(\mathbf{t},\mathbf{v})\in\mathbb{T}_{3N}$, preserves a minimum
pairwise distance of
$$\alpha^{\neq 0}_{\text{tot}}=\min\left(\alpha, \frac{L}{N^2}\right).$$

{\noindent \bf Case $\alpha=0$.} We define the subset of indices
$\mathcal{I}_f = \{l\in\setind; \exists i\in\setind, i\neq l,
x^{(f)}_{i,1}= x^{(f)}_{l,1}\}$, the equivalence classes
$[l] = \{i\in\mathcal{I}_f; x^{(f)}_{i,1}= x^{(f)}_{l,1}\}$ and the
bijections $\nu_l:[l]\to \llbracket 0, \text{Card}(l)-1\rrbracket$ so that
$ \nu_l(i)>\nu_l(j)$ if
$i>j$. We also now consider the quantity
$$\alpha^{\neq 0}=\min\left\{\min\left(|x_{i,1}^{(f)}-x^{(f)}_{j,1}|, L - |x_{i,1}^{(f)}-x^{(f)}_{j,1}|\right); i\in\setind\setminus\mathcal{I}_f, j\in\setind, i\neq j \right\} + \1_{\{\setind\}}(\mathcal{I}_f)L.$$
From the configuration $x^{(\mathsf{x})}$, we update the y-coordinates
of the spheres $i\in\setind\setminus\mathcal{I}_f$ to their final
values as,
   \begin{equation}
     \mathbf{t}_{\mathsf{y}}^{\neq 0} = \left(\left(x^{(f)}_{i,1} - \tilde{x}^{(0)}_{i,1}\right)\mod L\right)_{i\in\setind\setminus\mathcal{I}_f}, \;
      \mathbf{v}_{\mathsf{y}}^{\neq 0} = ((u_{\mathsf{y}},i))_{i\in\setind\setminus\mathcal{I}_f},
     \label{eq:sets_for_line}
 \end{equation}
the corresponding flow
$\phi_{\mathbf{t}_{\mathsf{y}}^{\neq
    0}}^{\mathbf{v}_{\mathsf{y}}^{\neq 0}}$ being well-defined, as it
preserves a minimum pairwise distance of at least $L/N^2$.
We now update the y-coordinates of the spheres $i\in\mathcal{I}_f$ as,
   \begin{equation}
     \mathbf{t}_{\mathsf{y}}^{= 0} = \left(\left(\left(x^{(f)}_{i,1} - \tilde{x}^{(0)}_{i,1}\right)\mod L\right) - \frac{\alpha^{\neq 0}\nu_i(i)}{|[i]|} \right)_{i\in\mathcal{I}_f}, \;
      \mathbf{v}_{\mathsf{y}}^{= 0} = ((u_{\mathsf{y}},i))_{i\in\mathcal{I}_f},
     \label{eq:sets_for_line}
 \end{equation}
the corresponding flow
$\phi_{\mathbf{t}_{\mathsf{y}}^{=
    0}}^{\mathbf{v}_{\mathsf{y}}^{= 0}}$ being well-defined, as it
preserves a minimum pairwise distance of at least  $L/N^2$.

Now, the x-coordinates of all the spheres
are updated to their final values by considering the sequences,
  \begin{equation}
    \mathbf{t}_{\mathsf{f,x}} = \left(\left(x^{(f)}_{i,0} - x^{(\mathsf{x})}_{i,0}\right)\mod L\right)_{i=1}^{N}, \;
     \mathbf{v}_{\mathsf{f,x}} = \left((u_{\mathsf{x}},i )\right)_{i=1}^{N},
    \label{eq:sets_for_line}
\end{equation}
and the corresponding flow
$\phi_{\mathbf{t}_{\mathsf{f,x}}}^{\mathbf{v}_{\mathsf{f,x}}}$
exists as it preserves a minimum pairwise distance of at least
$\alpha^{\neq 0}/\max_i|[i]|$.

Eventually, the y-coordinates of the spheres $i\in\mathcal{I}_f$ are
also updated to their final values by considering the sequences, 
  \begin{equation}
    \mathbf{t}_{\mathsf{f,y}} = \left(\frac{\alpha^{\neq 0}\nu_i(i)}{|[i]|} \right)_{i\in\mathcal{I}_f}, \;
     \mathbf{v}_{\mathsf{f,y}} = \left((u_{\mathsf{y}},i \right)_{i\in\mathcal{I}_f}, v^{(f)}),
    \label{eq:sets_for_line}
\end{equation}
and the corresponding flow
$\phi_{\mathbf{t}_{\mathsf{f,y}}}^{\mathbf{v}_{\mathsf{f,y}}}$
exists as it preserves a minimum pairwise distance of at least

$$\alpha_{\mathcal{I}_f}=\min_{i\in\mathcal{I}_f}\left\{\min_{\substack{l\in[i]\\l\neq i}}\left(\min\left(|x_{i,0}^{(f)}-x^{(f)}_{j,0}|, L - |x_{i,0}^{(f)}-x^{(f)}_{j,0}|\right)\right)\right\}>0.$$

Eventually, the total composite flow
$\phi_{\mathbf{t}}^{\mathbf{v}}=\phi_{\mathbf{t}_{\mathsf{f,y}}}^{\mathbf{v}_{\mathsf{f,y}}}\circ\phi_{\mathbf{t}_{\mathsf{f,x}}}^{\mathbf{v}_{\mathsf{f,x}}}\circ\phi_{\mathbf{t}_{\mathsf{y}}^{=0}}^{\mathbf{v}_{\mathsf{y}}^{y=0}}\circ\phi_{\mathbf{t}_{\mathsf{y}}^{y\neq
    0}}^{\mathbf{v}_{\mathsf{y}}^{\neq
    0}}\circ\phi_{\mathbf{t}_{\mathsf{x}}^{\mathcal{I}_0}}^{\mathbf{v}^{\mathcal{I}_0}_{\mathsf{x}}}\circ\phi_{\mathbf{t}_{\mathsf{x}}}^{\mathbf{v}_{\mathsf{x}}}$,
$(\mathbf{t},\mathbf{v})\in\mathbb{T}_{3N+|I_f|}$,
preserves a minimum pairwise distance of
$$\alpha^{=0}_{\text{tot}}=\min\left(\alpha_{\mathcal{I}_f},\frac{\alpha^{\neq 0}}{\max_i|[i]|}, \frac{L}{N^2}\right).$$

Thus, this procedure shows how to obtain a lower bound of the
reachability probability which depends only on the travel times
between initial and final positions, impacting the probability to get
the target number of refreshment events, ruled by a homogeneous
Poisson process. In future works, it could be interesting to build on
this alternative procedure to obtain a coupling strategy.

\section*{Acknowledgments}
 A.M. is grateful for the
support received from the Agence Nationale de la Recherche (ANR) of
the French government through the program "Investissements d'Avenir"
(16-IDEX-0001 CAP 20-25). A.G. is supported by the French ANR under
the grant ANR-17-CE40-0030 (project \emph{EFI}) and the Institut
Universtaire de France. M.M. acknowledges the support of the
French ANR under the grant ANR-20-CE46-0007 (\emph{SuSa} project).


\begin{thebibliography}{10}

\bibitem{Metropolis_1953}
Nicholas Metropolis, Arianna~W Rosenbluth, Marshall~N Rosenbluth, Augusta~H
  Teller, and Edward Teller.
\newblock Equation of state calculations by fast computing machines.
\newblock {\em The journal of chemical physics}, 21(6):1087--1092, 1953.

\bibitem{Alder_1962}
BJ~Alder and TE~Wainwright.
\newblock Phase transition in elastic disks.
\newblock {\em Physical Review}, 127(2):359, 1962.

\bibitem{Jaster_1999}
Andreas Jaster.
\newblock An improved metropolis algorithm for hard core systems.
\newblock {\em Physica A: Statistical Mechanics and its Applications},
  264(1-2):134--141, 1999.

\bibitem{Bernard_2011}
Etienne~P Bernard and Werner Krauth.
\newblock Two-step melting in two dimensions: first-order liquid-hexatic
  transition.
\newblock {\em Physical review letters}, 107(15):155704, 2011.

\bibitem{Diaconis_2011}
Persi Diaconis, Gilles Lebeau, and Laurent Michel.
\newblock {Geometric analysis for the metropolis algorithm on Lipschitz
  domains}.
\newblock {\em {Inventiones Mathematicae}}, 185(2):239--281, August 2011.

\bibitem{Janke_2002}
Wolfhard Janke.
\newblock Statistical analysis of simulations: Data correlations and error
  estimation.
\newblock {\em Quantum Simulations of Complex Many-Body Systems: From Theory to
  Algorithms}, 10:423--445, 2002.

\bibitem{Levin_2017}
David~A Levin and Yuval Peres.
\newblock {\em Markov chains and mixing times}, volume 107.
\newblock American Mathematical Soc., 2017.

\bibitem{Hohenberg_1977}
P.~C. Hohenberg and B.~I. Halperin.
\newblock Theory of dynamic critical phenomena.
\newblock {\em Rev. Mod. Phys.}, 49:435--479, Jul 1977.

\bibitem{Swendsen_1987}
Robert~H Swendsen and Jian-Sheng Wang.
\newblock Nonuniversal critical dynamics in monte carlo simulations.
\newblock {\em Physical review letters}, 58(2):86, 1987.

\bibitem{Wolff_1989}
Ulli Wolff.
\newblock Collective monte carlo updating for spin systems.
\newblock {\em Physical Review Letters}, 62(4):361, 1989.

\bibitem{Turitsyn_2011}
Konstantin~S Turitsyn, Michael Chertkov, and Marija Vucelja.
\newblock Irreversible monte carlo algorithms for efficient sampling.
\newblock {\em Physica D: Nonlinear Phenomena}, 240(4-5):410--414, 2011.

\bibitem{Peters_2012}
Elias~AJF Peters et~al.
\newblock Rejection-free monte carlo sampling for general potentials.
\newblock {\em Physical Review E}, 85(2):026703, 2012.

\bibitem{Bernard_2009}
Etienne~P. Bernard, Werner Krauth, and David~B. Wilson.
\newblock Event-chain monte carlo algorithms for hard-sphere systems.
\newblock {\em Phys. Rev. E}, 80:056704, Nov 2009.

\bibitem{Michel_2014}
Manon Michel, Sebastian~C. Kapfer, and Werner Krauth.
\newblock Generalized event-chain monte carlo: Constructing rejection-free
  global-balance algorithms from infinitesimal steps.
\newblock {\em The Journal of Chemical Physics}, 140(5):054116, 2014.

\bibitem{Widom_1987}
Michael Widom, Katherine~J Strandburg, and Robert~H Swendsen.
\newblock Quasicrystal equilibrium state.
\newblock {\em Physical review letters}, 58(7):706, 1987.

\bibitem{Kapfer_2015}
Sebastian~C. Kapfer and Werner Krauth.
\newblock Two-dimensional melting: From liquid-hexatic coexistence to
  continuous transitions.
\newblock {\em Phys. Rev. Lett.}, 114:035702, Jan 2015.

\bibitem{Kampmann_2015}
Tobias~A Kampmann, Horst-Holger Boltz, and Jan Kierfeld.
\newblock Monte carlo simulation of dense polymer melts using event chain
  algorithms.
\newblock {\em The Journal of chemical physics}, 143(4):044105, 2015.

\bibitem{Michel_2015}
Manon Michel, Johannes Mayer, and Werner Krauth.
\newblock Event-chain monte carlo for classical continuous spin models.
\newblock {\em EPL (Europhysics Letters)}, 112(2):20003, 2015.

\bibitem{Nishikawa_2015}
Yoshihiko Nishikawa, Manon Michel, Werner Krauth, and Koji Hukushima.
\newblock Event-chain algorithm for the heisenberg model: Evidence for
  z$\simeq$1 dynamic scaling.
\newblock {\em Physical Review E}, 92(6):063306, 2015.

\bibitem{Harland_2017}
Julian Harland, Manon Michel, Tobias~A Kampmann, and Jan Kierfeld.
\newblock Event-chain monte carlo algorithms for three-and many-particle
  interactions.
\newblock {\em EPL (Europhysics Letters)}, 117(3):30001, 2017.

\bibitem{Michel_2020}
Manon Michel, Alain Durmus, and Stéphane Sénécal.
\newblock Forward event-chain monte carlo: Fast sampling by randomness control
  in irreversible markov chains.
\newblock {\em Journal of Computational and Graphical Statistics},
  29(4):689--702, 2020.

\bibitem{Klement_2019}
Marco Klement and Michael Engel.
\newblock Efficient equilibration of hard spheres with newtonian event chains.
\newblock {\em The Journal of chemical physics}, 150(17):174108, 2019.

\bibitem{Hoellmer_2021}
Philipp Hoellmer, Nicolas Noirault, Botao Li, AC~Maggs, and Werner Krauth.
\newblock Sparse hard-disk packings and local markov chains.
\newblock {\em arXiv preprint arXiv:2109.13343}, 2021.

\bibitem{Bouchard_2018}
Alexandre Bouchard-C{\^o}t{\'e}, Sebastian~J Vollmer, and Arnaud Doucet.
\newblock The bouncy particle sampler: A nonreversible rejection-free markov
  chain monte carlo method.
\newblock {\em Journal of the American Statistical Association},
  113(522):855--867, 2018.

\bibitem{Bierkens_2019_zz}
Joris Bierkens, Paul Fearnhead, and Gareth Roberts.
\newblock The zig-zag process and super-efficient sampling for {B}ayesian
  analysis of big data.
\newblock {\em Ann. Statist.}, 47(3):1288--1320, 2019.

\bibitem{Davis_1984}
Mark~HA Davis.
\newblock Piecewise-deterministic markov processes: a general class of
  non-diffusion stochastic models.
\newblock {\em Journal of the Royal Statistical Society: Series B
  (Methodological)}, 46(3):353--376, 1984.

\bibitem{Davis_1993}
M.~H.~A. Davis.
\newblock {\em Markov models and optimization}, volume~49 of {\em Monographs on
  Statistics and Applied Probability}.
\newblock Chapman \& Hall, London, 1993.

\bibitem{Radulescu_2007}
Ovidiu Radulescu, Aur{\'e}lie Muller, and Alina Crudu.
\newblock Th{\'e}or{\`e}mes limites pour les processus de markov {\`a} sauts.
\newblock {\em Tech. Sci. Informatiques}, 26(3-4):443--469, 2007.

\bibitem{Yvinec_2015}
Romain Yvinec.
\newblock {Piecewise deterministic Markov processes, applications in biology}.
\newblock Lecture, 2015.

\bibitem{M2015}
Florent Malrieu.
\newblock Some simple but challenging {M}arkov processes.
\newblock {\em Ann. Fac. Sci. Toulouse Math. (6)}, 24(4):857--883, 2015.

\bibitem{Durmus_2018}
Alain Durmus, Arnaud Guillin, and Pierre Monmarch{\'e}.
\newblock {Piecewise Deterministic Markov Processes and their invariant
  measure}.
\newblock To appear in {\it Annales de l'Institut Henri Poincar\'e}, July 2018.

\bibitem{Fearnhead_2018}
Paul Fearnhead, Joris Bierkens, Murray Pollock, and Gareth~O. Roberts.
\newblock Piecewise deterministic {M}arkov processes for continuous-time
  {M}onte {C}arlo.
\newblock {\em Statist. Sci.}, 33(3):386--412, 2018.

\bibitem{Meyn_1993}
Sean~P Meyn and Richard~L Tweedie.
\newblock Stability of markovian processes ii: Continuous-time processes and
  sampled chains.
\newblock {\em Advances in Applied Probability}, 25(3):487--517, 1993.

\bibitem{Tweedie_1994}
R.~L. Tweedie.
\newblock Topological conditions enabling use of harris methods in discrete and
  continuous time.
\newblock {\em Acta Applicandae Mathematicae}, 34(1-2):175--188, February 1994.

\bibitem{Meyn_2012}
Sean~P Meyn and Richard~L Tweedie.
\newblock {\em Markov chains and stochastic stability}.
\newblock Springer Science \& Business Media, 2012.

\bibitem{Bakhtin_2012}
Yuri Bakhtin and Tobias Hurth.
\newblock Invariant densities for dynamical systems with random switching.
\newblock {\em Nonlinearity}, 25(10):2937--2952, October 2012.

\bibitem{Benaim_2015}
Michel Benaïm, Stéphane Le~Borgne, Florent Malrieu, and Pierre-André Zitt.
\newblock Qualitative properties of certain piecewise deterministic {Markov}
  processes.
\newblock {\em Annales de l'Institut Henri Poincaré, Probabilités et
  Statistiques}, 51(3), August 2015.

\bibitem{Bierkens_2019}
Joris Bierkens, Gareth~O Roberts, Pierre-Andr{\'e} Zitt, et~al.
\newblock Ergodicity of the zigzag process.
\newblock {\em The Annals of Applied Probability}, 29(4):2266--2301, 2019.

\bibitem{Hastings_1970}
W.~K. Hastings.
\newblock {Monte Carlo sampling methods using Markov chains and their
  applications}.
\newblock {\em Biometrika}, 57(1):97--109, 04 1970.

\bibitem{Diaconis_2000}
Persi Diaconis, Susan Holmes, and Radford~M Neal.
\newblock Analysis of a nonreversible markov chain sampler.
\newblock {\em Annals of Applied Probability}, pages 726--752, 2000.

\bibitem{Chevallier2021}
Augustin Chevallier, Sam Power, Andi~Q. Wang, and Paul Fearnhead.
\newblock Pdmp monte carlo methods for piecewise-smooth densities.
\newblock {\em arXiv preprint arXiv:2111.05859}, 2021.

\bibitem{H2021}
Peter Holderrieth.
\newblock Cores for piecewise-deterministic {M}arkov processes used in {M}arkov
  chain {M}onte {C}arlo.
\newblock {\em Electron. Commun. Probab.}, 26:Paper No. 57, 12, 2021.

\bibitem{Durmus_2020}
Alain Durmus, Arnaud Guillin, and Pierre Monmarch\'{e}.
\newblock Geometric ergodicity of the bouncy particle sampler.
\newblock {\em Ann. Appl. Probab.}, 30(5):2069--2098, 2020.

\bibitem{Lu_2018}
Yulong Lu and Jonathan~C. Mattingly.
\newblock Geometric ergodicity of {L}angevin dynamics with {C}oulomb
  interactions.
\newblock {\em Nonlinearity}, 33(2):675--699, 2020.

\bibitem{Azema_1967}
J~Azema, M~Kaplan-Duflo, and D~Revuz.
\newblock Mesure invariante sur les classes r{\'e}currentes des processus de
  markov.
\newblock {\em Zeitschrift f{\"u}r Wahrscheinlichkeitstheorie und Verwandte
  Gebiete}, 8(3):157--181, 1967.

\bibitem{Getoor_1980}
Ronald~K Getoor.
\newblock Transience and recurrence of markov processes.
\newblock In {\em S{\'e}minaire de Probabilit{\'e}s XIV 1978/79}, pages
  397--409. Springer, 1980.

\bibitem{Tuominen_1979}
Pekka Tuominen and Richard~L Tweedie.
\newblock The recurrence structure of general markov processes.
\newblock {\em Proceedings of the London Mathematical Society}, 3(3):554--576,
  1979.

\end{thebibliography}
\end{document}